\documentclass[nofigure]{pasj01}
\draft
\usepackage{graphicx}

\begin{document}
\SetRunningHead{T.\ Ueta et al.}{AKARI FIS Flux Correction for Compact Extended Sources}
\title{Surface Brightness Correction for Compact Extended Sources Observed by the AKARI Far-Infrared Surveyor (FIS) in the Slow-Scan Mode}

\Received{July 28, 2016}
\Accepted{TBD}

\author{%
Toshiya \textsc{Ueta,}\altaffilmark{1}
Rachael L. \textsc{Tomasino,}\altaffilmark{1}
Satoshi \textsc{Takita,}\altaffilmark{2}
Hideyuki \textsc{Izumiura,}\altaffilmark{3,4}
Mai \textsc{Shirahata,}\altaffilmark{2}
Andrew \textsc{Fullard,}\altaffilmark{1}
Issei \textsc{Yamamura,}\altaffilmark{2,5}
and
Shuji \textsc{Matsuura}\altaffilmark{6}}

\altaffiltext{1}{Department of Physics and Astronomy, University of Denver, Denver, CO 80208, USA}
\altaffiltext{2}{Institute of Space and Astronautical Science, Japan Aerospace Exploration Agency, Yoshinodai 3-1-1 , Chuo-ku, Sagamihara, Kanagawa, 252-5210, Japan}
\altaffiltext{3}{Okayama Astrophysical Observatory, National Astronomical Observatory of Japan, National Institutes of Natural Sciences, Honjo 3037-5, Kamogata, Asakuchi, Okayama 719-0232, Japan}
\altaffiltext{4}{National Institutes of Natural Science, National Astronomical Observatory of Japan (NAOJ), Osawa 2-21-1, Mitaka, Tokyo, 181-8588, Japan}
\altaffiltext{5}{Department of Space and Astronautical Science, School of Physical Sciences, SOKENDAI, Yoshino-dai 3-1-1, Chuo-ku, Sagamihara, kanagawa, 252-5210, Japan}
\altaffiltext{6}{Kwansei Gakuin University, Department of Physics, School of Science and Technology, Gakuen 2-1, Sanda, Hyogo, 669-1337, Japan}


\maketitle

\begin{abstract}
We present a general surface brightness correction method for compact extended sources imaged in the slow-scan pointed observation mode of the Far-Infrared Surveyor (FIS) aboard the {\sl AKARI\/} Infrared Astronomical Satellite.
Our method recovers correct surface brightness distribution maps by re-scaling archived raw FIS maps using the surface-brightness-dependent inverse FIS response function.
The flux of a target source is then automatically corrected for as the simple sum of surface brightnesses within the adopted contour encircling the perimeter of the target (i.e., contour photometry). 
This correction method is contrasted to the previous aperture photometry method for point sources, which directly corrects for the target flux with a flux-dependent scaling law.
The new surface brightness correction scheme is applicable to objects of any shape from unresolved point sources to resolved extended objects, as long as the target is not deemed diffuse, i.e., the total extent of the target source does not exceed too much more than a single FIS scan width of $10^{\prime}$. 
The new correction method takes advantage of the well-defined shape (i.e., the scale invariance) of the point-spread function, which enables us to adopt a power-law FIS response function.
We analyze the point-source photometric calibrator data using the FIS {\sl AKARI\/} Slow-scan Tool (FAST) and constrained the parameters of the adopted power-law FIS response function.
We conclude that the photometric accuracy of the new correction method is better than 10\% error based on comparisons with the expected fluxes of the photometric calibrators and that resulting fluxes without the present correction method can lead up to 230\% overestimates or down to 50\% underestimates.
\end{abstract}

\section{Introduction}

The {\sl AKARI\/} Infrared Astronomical Satellite ({\sl AKARI\/}; \cite{akari}) is the Japanese infrared space mission launched on 2006 February 21 (UT).
The mission goals of {\sl AKARI\/} are to 
(1) perform a high-spatial resolution all-sky survey in 6 infrared bands from 9 to 160\,$\micron$ for the first time since the Infrared Astronomical Satellite ({\sl IRAS\/}; \cite{Neugebauer_1984}) and 
(2) conduct pointed observations of specific targets by spending roughly 20\% of the on-orbit time to obtain deeper images and spectroscopic data (from 2 to 180\,$\micron$).
{\sl AKARI\/} carried out its 550-day cryogen mission until it exhausted liquid Helium on 2007 August 26, and continued its post-cryogen mission in the near-infrared until the satellite was finally turned off on 2011 November 24.

The Far-Infrared Surveyor (FIS; \cite{Kawada_2007}) is one of the two instruments on-board {\sl AKARI\/}, covering the wavelength range of 50 to 180$\,\mu$m with two sets of Ge:Ga arrays, the Short-Wavelength (SW; \cite{Fujiwara_2003}) and Long-Wavelength (LW; \cite{Doi_2002}) detectors in the N60 (50 to 80\,$\micron$) and WIDE-S (60 to 110$ \mu$m) bands and the WIDE-L (110 to 180$\,\mu$m) and N160 (140 to 180$ \mu$m) bands, respectively. 
During all-sky survey observations, the sky was swept at $3\farcm6$\,s$^{-1}$ covering more than 98\,\% of the entire sky (the all-sky scan mode; \cite{Doi_2015,Takita_2015}).
During pointed observations, on the other hand, intended targets were scanned at $8^{\prime\prime}$\,s$^{-1}$ or $15^{\prime\prime}$\,s$^{-1}$ to achieve 1 to 2 orders of magnitude better sensitivity than the all-sky survey observations (the slow scan mode; \cite{Shirahata_2009}).

The absolute surface-brightness calibration of the FIS instrument was done through 
(1) pre-launch laboratory measurements of a blackbody source which indicated a 5$\%$ accuracy, and
(2) on-orbit comparisons between measurements of infrared cirrus regions without significant small-scale structures made by FIS and the DIRBE instrument on-board the {\sl COBE} satellite \citep{m11}. 
Hence, the FIS data (both all-sky and slow scan data) presently archived\footnote{In the Data ARchives and Transmission System (DARTS), http://www.darts.isas.ac.jp/astro/akari/.} 
should give correct surface brightnesses of diffuse background emission.
However, when aperture photometry was performed for a set of infrared flux calibrators detected in the calibrated FIS slow-scan maps, the resulting fluxes came out to be roughly 40\% less than expected \citep{Shirahata_2009}, which is attributed to the slow transient response of the Ge:Ga detectors (e.g., \cite{Kaneda_2002})\footnote{Note that the absolute flux calibration for the {\sl AKARI\/}/FIS All-Sky Bright Source Catalogue was done differently by directly comparing the model flux of reference objects and the time-series detector signal readouts \citep{Yamamura_2010}.}.

To alleviate this issue, already devised were methods of flux correction specifically aimed at point sources observed with FIS in the slow-scan mode \citep{Shirahata_2009}.
These point-source flux correction methods are based on the premise that the shape of the point-spread function (PSF) is well-defined: the point-source flux within an idealized infinite aperture can always be recovered by scaling the raw flux measured within a finite aperture (which usually covers at least the brightest PSF core) by an appropriate aperture correction factor.
However, such a correction method would not work in general for objects that are neither point-like nor diffuse, because the surface brightness distribution of such objects is not known {\sl a priori} (e.g., circumstellar shells and nearby galaxies, which we refer to collectively as ``compact extended sources'' here, e.g., \cite{Kaneda_2007,Hirashita_2008,Ueta_2008,Kaneda_2010,Cox_2011,Izumiura_2011}).
Hence, there is a definite need for a more versatile surface brightness correction scheme aimed at compact extended sources observed with {\sl AKARI\/}/FIS in the slow-scan pointed observation mode.

Therefore, we present a new, more general surface brightness (and hence flux) correction scheme for far-IR images of compact extended sources (defined to be less extended than 10$^{\prime}$, which corresponds to the single-scan angular width of an FIS map; \cite{Kawada_2007}) observed with {\sl AKARI\/}/FIS in the slow-scan pointed observation mode.
Below, we formulate a new correction scheme by adopting the power-law FIS response function under the premise that the PSF shape remains unchanged irrespective of the source brightness (\S\,\ref{scheme}),
discuss the re-analysis of the point-source flux calibrator data (\S\,\ref{data}),
derive the parameters of the adopted FIS response function and assess the quality of the new surface brightness correction scheme (\S\,\ref{fluxcalibration}),
before summarizing the entire analysis (\S\,\ref{summary}).
We emphasize again that the new surface brightness correction scheme presented here is intended specifically for slow-scan pointed observations and not for all-sky scan observations.
Presently FIS slow-scan data are being processed into images by the {\sl AKARI\/} team: all valid observations in the SW band (N60 and WIDE-S) and good-quality observations in the LW band (WIDE-L and N160), along with relevant softwares, will be made publicly available\footnote{Check back http://www.ir.isas.jaxa.jp/AKARI/Observation/ for the latest and updates.} (Takita et al.\ {\sl in preparation}).
Those who wish to use {\sl AKARI\/}/FIS maps (especially taken in the same observing mode as the present work) for their science that involves objects that are extended (but more compact than about $10^{\prime}$) are encouraged to adopt this correction scheme to obtain correct surface brightness distributions of the target sources.

\section{\label{scheme}%
New Surface Brightness Correction Scheme for Compact Extended Source}

\citet{Shirahata_2009} performed in-orbit performance verification of the FIS slow-scan pointed observation mode and found that
(1) the sensitivity of FIS slow-scan maps to point sources was lower than that to diffuse background emission,
resulting in fluxes that were about 20\% to 80\% of the expected values, and
(2) the observed sensitivity variation was apparently dependent on the strength of incoming signals falling onto the FIS arrays.
These findings were based on comparisons between fluxes of the infrared photometric reference sources measured via aperture photometry and the model predictions.
Such reduced sensitivity to brighter point sources was attributed to the delay in response to the incoming flux known to occur with Ge:Ga detectors (hence, ``slow" transient response; \cite{isocam,Kaneda_2002,mips}).
The slow transient response in scan observations of bright objects manifests itself as underestimated peak surface brightnesses, resulting in lower flux values.

Obviously, objects that are neither point-like nor diffuse (i.e., compact extended sources) are also expected to be influenced by the same sensitivity variation due to the slow transient response.
However, the extent to which fluxes of these compact extended sources would be affected is not immediately obvious, especially because the above assessments for point-source fluxes were based on aperture photometry, which cannot be applied directly to compact extended sources whose shape is not necessarily point-like.
The only practical way to measure flux from compact extended sources is to define the outer perimeter by some sort of surface brightness thresholding and take the sum of the surface brightnesses within the perimeter (i.e., contour photometry).
Thus, if we were to correct for the flux of such compact extended sources measured by contour photometry, the only way to do so is to re-scale the surface brightnesses of the archived slow-scan maps {\sl before} performing contour photometry.

To proceed, let us consider the empirical FIS response function, $\mathcal{R}$, such that
$S_{ij, {\rm FIS}} = \mathcal{R} (S_{ij})$,
where $S_{ij}$ is the incoming surface brightness distribution of the mapped region of the sky in the far-infrared and $S_{ij, {\rm FIS}}$ is the measured surface brightness distribution seen in the resulting archived FIS map.
The subscripts $i$ and $j$ refer to individual pixels.
If we can determine the inverse FIS response function $\mathcal{R}^{-1}$, we would then re-scale the archived FIS map by this function $\mathcal{R}^{-1}$ to recover $S_{ij} = \mathcal{R}^{-1}(S_{ij, {\rm FIS}})$ and obtain the ``correct" flux of a target source by performing contour photometry on the re-scaled map, $S_{ij}$.

Here, we reiterate with the fact that the shape of the point-spread function (PSF) in the FIS slow-scan maps is determined to be stable irrespective of the source brightness and color (\cite{Shirahata_2009}; especially their Figures 3 and 4).
Consider the surface brightness distribution of a PSF reference object incident on the FIS arrays, $S_{ij}^{\rm PSF} = I_{0}f_{ij}$, where $I_{0}$ and $f_{ij}$ are respectively the peak intensity and normalized surface brightness profile of the {\sl AKARI\/}/FIS PSF reference object at this instance.
It then immediately follows that the flux of this PSF reference is $F_{\rm PSF} = \sum S_{ij}^{\rm PSF} = I_{0} \sum f_{ij}$, where the summation symbol represents the essence of aperture/contour photometry.
This $F_{\rm PSF}$ is the correct source flux out of the re-scaled FIS map, $S_{ij}$.
This object would appear in the archived FIS map as $S_{ij, {\rm FIS}}^{\rm PSF} = I_{1} f^{\prime}_{ij}$, where $I_{1}$ and $f^{\prime}_{ij}$ are the corresponding peak intensity and normalized surface brightness profile of the {\sl AKARI\/}/FIS PSF in the archived FIS map.
The raw flux out of the archived FIS maps would then be $F_{\rm PSF,FIS} = \sum S_{ij, {\rm FIS}}^{\rm PSF} = I_{1} \sum f^{\prime}_{ij}$.
According to the previous assessments by \citet{Shirahata_2009}, $F_{\rm PSF,FIS}$ is underestimated by 20\% to 80\% with respect to $F_{\rm PSF}$.

The observed PSF shape uniformity in this formulation means that the empirical normalized surface brightness profile of a PSF, $f^{\prime}_{ij}$, is always the same in FIS maps, and that the raw PSF flux, $F_{\rm PSF,FIS}$, is uniquely determined as soon as the observed PSF peak intensity, $I_{1}$, is specified.
Correspondingly, the correct PSF flux, $F_{\rm PSF}$, is uniquely determined as soon as the correct PSF peak intensity, $I_{0}$, is specified, because the correct normalized surface brightness profile of a PSF, $f_{ij}$, is always the same.
Of course, $f_{ij}$ and $f^{\prime}_{ij}$ are related via the FIS response function, $\mathcal{R}$, via $f^{\prime}_{ij} = \mathcal{R}(f_{ij})$ as $\mathcal{R}(I_{0})=I_{1}$.
At the same time, neither $f_{ij}$ nor $f^{\prime}_{ij}$ would vary with the source brightness and color.
Hence, the pixel-wise scaling between $f_{ij}$ and $f^{\prime}_{ij}$ must always be fixed.

Mathematically speaking, the observed PSF shape uniformity requires that the empirical FIS response function, $\mathcal{R}$, is scale invariant.
Because a power-law function is scale-invariant, it is appropriate to assume that the empirical FIS response function, $\mathcal{R}$, is a power-law function, i.e., 
\begin{equation}\label{fisresponse}
S_{ij, {\rm FIS}} = \mathcal{R}(S_{ij}) = c S_{ij}^n,
\end{equation}
where $n$ and $c$ are, respectively, the power-law index and the scaling coefficient of the adopted power-law FIS response function\footnote{The most ideal situation is, of course, when the detector response is linear ($n=1$) and the scaling is unity ($c=1$).}.
Hence, the re-scaling of the archived FIS maps to regain the ``correct" surface brightness distribution falling onto the FIS detectors can be done by 
\begin{equation}\label{rescaling}
S_{ij} = \mathcal{R}^{-1}(S_{ij, {\rm FIS}}) = (S_{ij, {\rm FIS}}/c)^{1/n}.
\end{equation}
Therefore, a flux correction for compact extended sources is reduced to determining the $(n, c)$ parameters that properly re-scale the {\sl AKARI}/FIS slow-scan maps of the target compact extended sources.
Then, the corrected flux of a target source, $F_{\rm SRC}$, would be
\begin{equation}\label{correctflux}
F_{\rm SRC} = \sum S_{ij} = \sum [(S_{ij, {\rm FIS}}/c)^{1/n}] = (1/c)^{1/n} \sum [(S_{ij, {\rm FIS}})^{1/n}].
\end{equation}
Note that the last sum must be done {\sl after} the archived FIS map is raised by the $1/n$ power, because in general $\sum [(S_{ij, {\rm FIS}})^{1/n}] \ne  (\sum S_{ij, {\rm FIS}})^{1/n} = (F_{\rm SRC,FIS})^{1/n}$, where $F_{\rm SRC,FIS}$ is the raw source flux measured straight out of the archived FIS map.
This is also why flux correction for compact extended sources must be performed by re-scaling the surface brightness map of a target before obtaining the flux.
Last but not least, this formulation is general enough that the scheme would work to recalibrate the surface brightness maps that suffer from similar reduction of sensitivity depending on the compactness of the target sources.

\section{\label{data}%
Data Preparation}

\subsection{PSF/Photometric Reference Objects}

The original flux correction scheme for point sources detected in the {\sl AKARI\/}/FIS slow-scan maps \citep{Shirahata_2009} was established based on observations of well-defined infrared PSF/photometric standard sources
(stars: \cite{cohen1999}, \yearcite{cohen2003a}, \yearcite{cohen2003b}; 
asteroids and planets: \cite{mullerlagerros1999}, \yearcite{mullerlagerros2002}; \cite{moreno_1998}). 
All observations were carried out using the Astronomical Observation Template (AOT) FIS01 scan sequence, which consists of two pairs of forward and reverse linear scans with a cross-scan spacing between the two scans. 
The scan speed was 8$^{\prime\prime}$\,s$^{-1}$ for most cases to achieve a high signal-to-noise ratio, while a few observations were performed with the 15$^{\prime\prime}$\,s$^{-1}$ in order to assess the influence of the scan speed on the data quality. 
The detector reset interval (0.5, 1.0, or 2.0 s) was chosen appropriately for each target to avoid saturation. 
Table\,\ref{obs_log} summarizes these reference objects used in the present analysis with their archival IDs, dates of observation, AOT parameters used, and expected (i.e, model prediction) fluxes with uncertainties.
For the present analysis, we used data from 10 stars (12 observations) and 10 solar system objects (16 observations).
Note that the expected fluxes listed in Table\,\ref{obs_log} are corrected for assuming a flat spectrum (i.e., $\nu F_{\nu} = \lambda F_{\lambda} = \textrm{const.}$) at each of the FIS bands.
Hence, there is no need to perform color correction in assessing the present analysis, especially when we compare cross-platform fluxes as long as a flat spectrum is assumed.

~\\
{\bf [Place Table\,\ref{obs_log} here]}\\


\subsection{\label{datareduction}%
Data Reduction with FAST}

The {\sl AKARI\/} slow-scan mapping data in the archive are stored in the time-series data (TSD) format. 
Therefore, they need to be processed into coadded maps with a map-making software.
We used the second-generation data reduction package, FIS AKARI Slow-scan Tool (FAST; \cite{Ikeda_2012}), which allows more flexible and thorough corrections of the TSD than the first-generation pipeline software, FIS Slow-Scan data-analysis Toolkit (SS-Tool: \cite{Matsuura_2007}) does. 
FAST is similar to SS-Tool, but allows for superior glitch and calibration lamp after-effect removal \citep{Suzuki_2008}. 
FAST also allows for more flexible selective exclusion of data that are affected by anomalies to produce less noisier final co-added maps.
Yet another difference is how photon energy is assumed to be distributed over the FIS arrays after each photon hit.
With SS-Tool photon energy is always assumed to get distributed uniformly over pixels within the beam, whereas with FAST other distribution profiles such as a Gaussian function are allowed for users to select.
The resulting FAST final co-added maps -- flux calibrated with respect to the diffuse cirrus emission -- were produced in the pixel scale of 8$^{\prime\prime}$\,pixel$^{-1}$ (corresponding to roughly 1/4 to 1/6 of the spatial resolution).  
Below FAST data reduction package is briefly described. 
Corresponding details of SS-Tool are found elsewhere (e.g., \cite{Matsuura_2007}).

With FAST, the raw TSD are first fed through the \textit{Green Box} pipeline, which processes and calibrates the TSD by 
(1) flagging bad data affected by dead and saturated pixels, reset anomalies, and other discontinuities, 
(2) converting the detector output from arbitrary data unit to volt, 
(3) correcting the detector signal to be proportional to the total incoming photon flux, 
(4) differentiating the data to represent the sky brightness, 
and 
(5) detecting and masking charged particle hit events/glitches
\citep{Verdugo_2007}.

In the AOT FIS01 scan sequence, calibration measurements are taken for a total of 5 times with the shutter closed before and after each of the four parallel scan legs \citep{Kawada_2007}.
The dark current and calibration lamp signals are monitored during each of these five calibration sequences to follow the time-varying instrument responsivity. 
On the other hand, self flat-field measurements, in which a flat frame is constructed to correct for the pixel-to-pixel detector responsivity variations, are made with the shutter opened by exposing the detector to the ``flat" sky during the calibration sequences at the beginning and end of the entire scan sequence (pre-cal and post-cal) while the telescope transitions between the all-sky survey mode and the Slow-Scan pointed observation mode \citep{Matsuura_2007}.

While SS-Tool is set to use all of the dark subtraction, detector responsivity time-variation correction, and the pre-cal flat-fielding, 
FAST allows users to determine whether or not particular calibration measurements are used in the data processing with the help of the GUI which visualizes the calibration measurements in the TSD \citep{ikeda_2012a}. 
This flexible selection of calibration measurements with FAST greatly enhances the effectiveness of the corrections and improves the resulting data quality, especially because it is now possible to eliminate calibration data that are compromised by anomalies.
We took advantage of this great capability of FAST to optimize the effectiveness of the calibration measurements.
In the final map-making, we oriented the images to align with the scan direction (i.e., the image $-y$ direction is the scan direction; also see Fig.\,3 of \cite{Kawada_2007}) and used the following options: the Earth-shine/stray-light removal, 200-sec median filtering, 5-$\sigma$ clipping, and Gaussian gridding convolution function of the beam (FWHM) size of $30^{\prime\prime}$ and $50^{\prime\prime}$ for the SW and LW bands, respectively (see \cite{Verdugo_2007} and \cite{ikeda_2012a} for more details).
The choice of the particular Gaussian beam size was dictated by the desire to keep general consistency with the all-sky scan maps \citep{Doi_2015,Takita_2015}.

\section{\label{fluxcalibration}%
Derivation of the Parameters of the Power-Law FIS Response Function}

\subsection{\label{invariance}Scale-Invariance of the {\sl AKARI\/}/FIS PSF}

As we outlined in \S\,\ref{scheme}, the scale invariance of the PSF shape permits us to adopt a power-law FIS response function, $\mathcal{R}$, in the form of $S_{ij, {\rm FIS}} = \mathcal{R}(S_{ij}) = c S_{ij}^n$ (eq.\,\ref{fisresponse}) and establish a scheme to recover $S_{ij}$ from $S_{ij, {\rm FIS}}$ by inverting this function as  $S_{ij} = (S_{ij, {\rm FIS}}/c)^{1/n}$ (eq.\,\ref{rescaling}).
Therefore, we need to verify the scale invariance of the {\sl AKARI\/}/FIS PSF.

Figure \ref{psfmad} shows the ``super-PSF" images in the four {\sl AKARI\/}/FIS bands made by taking the median of the normalized FAST-processed FIS maps of the PSF/photometric reference sources (using 24 and 18 sources for the SW and LW bands, respectively).
The dotted contours delineate the 3-$\sigma$ detection level in each of the 4 super-PSF images.
In essence, these super-PSF maps are the 2-D representations of $f^{\prime}_{ij}$ discussed in \S\,\ref{scheme}.
The SW band super-PSFs are fairly circular in the core, while the effects of the detector cross-talk signals along the long-axis of the FIS arrays are apparent (at $63\fdg5$ with respect to the image $-y$ axis, which is the scan direction).
The LW band PSFs are elongated along the short-axis of the FIS arrays, but suffer much less from the cross-talk signals (while the background is noisier). 

The 2-D single-peak gaussian fitting to the PSF core yielded the major-axis FWHM size (delineated by the dashed contours in Figure\,\ref{psfmad}) of 
$46\farcs7\pm6\farcs8$, 
$49\farcs2\pm6\farcs9$, 
$77\farcs9\pm6\farcs8$, and 
$71\farcs4\pm6\farcs4$, 
with the eccentricity of
0.10, 0.25, 0.48, and 0.54,
for the N60, WIDE-S, WIDE-L, and N160 bands, respectively.
The overall PSF structure is consistent with previously-reported in-flight performance \citep{Kawada_2007,Shirahata_2009}, confirming that FAST yields reasonable results.

~\\
{\bf [Place Figure\,\ref{psfmad} here]}\\

Also shown in Figure\,\ref{psfmad} are the median absolute deviation (MAD) maps for each band, made by taking the median of the absolute difference between each PSF image and the super-PSF image.
The MAD maps graphically represent how individual PSF images are statistically different from the corresponding super-PSF image at each pixel.
Within the region that registers more than 5-$\sigma$ (dashed contour in the PSF images), the median MADs intrinsic to the source emission are
$0.5\pm0.6\,\%$,
$0.8\pm0.4\,\%$,
$2.4\pm0.7\,\%$, and
$4.1\pm2.1\,\%$,
for the N60, WIDE-S, WIDE-L, and N160 bands, respectively.
These values indicate that the shape of the {\sl AKARI\/}/FIS PSF, $f^{\prime}_{ij}$, is identical more than 99\,\% in the SW bands and more than 95\,\% in the LW bands, i.e., the PSF shape is fairly uniform irrespective of the source brightnesses and object color/temperature (Tables\,\ref{obs_log} and \ref{FASTscaledphoto}). 
Hence, these super-PSF maps and their corresponding MAD maps confirm that $f^{\prime}_{ij}$ is uniform in all {\sl AKARI\/} bands, i.e., the {\sl AKARI\/}/FIS PSF is scale-invariant.
Thus, the scale invariance of the empirical FIS response function, $\mathcal{R}$, is duly warranted, and so is our adoptation of a power-law function for the FIS response function.
Here, we note in passing that the PSF uniformity is confirmed only within the parameter space explored, namely of the scan speed, reset interval, and cross-scan shift length, used to obtain the PSF reference data (see Table\,2 of \cite{Shirahata_2009}).

\subsection{\label{underestimateissue}%
Aperture Photometry vs.\ Contour Photometry}

Photometric measurements of point sources are most often made by the aperture photometry method.
With this method, the surface brightness of a point source is integrated within a finite (usually circular) aperture that covers the brightest PSF core. 
Then, this partial flux within the PSF core is scaled by an aperture correction factor to recover the total flux of the PSF under an idealized infinite aperture.
Hence, the aperture correction method is, by design, valid only for point sources.
This is because the method requires that the surface brightness profile of the source PSF is well-defined so that the aperture correction scaling factor does not vary in individual observations (i.e., the PSF shape is scale invariant; e.g., Table\,5 of \cite{Shirahata_2009}).
Moreover, the corrected flux for an infinite aperture can include the surface brightness component that is not explicitly detected by the detector.

Objects with extended emission do not usually follow a specific surface brightness profile.
Hence, flux measurements of extended sources cannot be made by the aperture correction method.
The only alternative to make photometric measurements of extended objects is the contour photometry method.
With this method, the surface brightness of a target object is spatially integrated over the {\sl entire} extent of the object that is defined by an object-specific contour at a certain thresholding signal-to-noise (S/N) level.
That is, the flux of the target is the sum of surface brightnesses within this finite bounding contour.
Thus, the contour photometry method would not allow for extrapolation of the surface brightness component that is not explicitly detected by the detector.

Therefore, to correct for the flux of compact extended sources with the contour photometry method, one must apply a correction to the surface brightness map of the source itself {\sl before} the corrected surface brightnesses are spatially integrated.
This is true especially when the detector sensitivity is dependent on the incoming surface brightness.
In case of FIS maps, this surface-brightness-dependent correction is indeed done via the inverse FIS response function (eq.\,\ref{rescaling}): the surface-brightness-corrected FIS maps, $S_{ij}$ (i.e., the real sky), are recovered by re-scaling the archived raw FIS maps, $S_{ij, {\rm FIS}}$, via $(S_{ij, {\rm FIS}}/c)^{1/n}$.
Hence, to establish the contour photometry method for compact extended sources detected in FIS maps, the $(n,c)$ scaling parameter pair must be constrained from observations of target sources of known fluxes, i.e., the photometric standards.
Note that this map re-scaling scheme is general enough that it would have to work for any source (point-source and compact extended source alike) as long as the source is not considered diffuse (in which case the archived maps are already absolutely calibrated).
However, as pointed out above, there is an intrinsic limitation for the contour photometry method:
fluxes obtained by this method are based solely on surface brightnesses that are detected within the photometric contour and there is no mechanism to recover the surface brightness component that may be present beyond the photometric contour or under the detection limit.

\subsection{\label{needs}%
Need for Surface Brightness Correction for FIS Maps}

In this section, we reconfirm the need for surface brightness correction for {\sl AKARI\/}/FIS maps when adopting the contour photometry method, by using PSF/photometric reference sources listed in Table\,\ref{obs_log}.
First, raw fluxes of these photometric references were measured by summing the uncorrected surface brightness pixel counts (in MJy\,sr$^{-1}$) within the photometric contour and converting this sum into flux (in Jy; with the pixel scale of the scan map, $8^{\prime\prime}$\,pixel$^{-1}$).
Here, the photometric contour was defined to be the 3-$\sigma$ detection level, where $\sigma$ is the S/N ratio measured in each of the PSF reference maps.
The background sky emission component whose spatial scale is roughly equal to or larger than the map size was already subtracted by median-filtering when scan maps were created at the end of the FAST pipeline processing. 

The uncorrected PSF fluxes measured in this way are bound to underestimate the true PSF fluxes because the contour photometry method can account for surface brightnesses only within the adopted finite 3-$\sigma$ contour, as opposed to the aperture photometry method in which surface brightnesses are accounted for in an idealized infinite aperture.
In general, the contour photometry method always underestimates source fluxes, and the degree of underestimation is more pronounced for dimmer objects for which the photometric contours become necessarily small.
Hence, for the present contour photometry, we emulated the uncorrected PSF fluxes under an idealized infinite aperture by appropriately scaling up the measured uncorrected PSF fluxes within the 3-$\sigma$ photometric contour.
The divisible scaling factor was determined as the ratio of the ``flux" of the normalized super-PSF map within the adopted 3-$\sigma$ contour (of the individual PSF map) to that within the 1-$\sigma$ contour (of the super-PSF map), which is the empirical equivalent to an infinite aperture.

~\\
{\bf [Place Figure\,\ref{rawphot} here]}\\

Figure\,\ref{rawphot} shows the brightness-dependent response of the FIS detector via the uncorrected-to-expected flux ratio as a function of the expected flux by comparing uncorrected PSF fluxes obtained with the expected model PSF fluxes (Table\,\ref{obs_log}).
The uncertainties were calculated by propagating both the instrumental error estimates and fluctuations in the residual sky emission.
The relationship can be fit by a power-law,
$(1.04 \pm 0.04) \times F_{\rm Jy}^{-0.10 \pm 0.01}$,
$(1.24 \pm 0.02) \times F_{\rm Jy}^{-0.11 \pm 0.01}$,
$(1.65 \pm 0.04) \times F_{\rm Jy}^{-0.16 \pm 0.01}$, and
$(0.54 \pm 0.02) \times F_{\rm Jy}^{-0.05 \pm 0.01}$,
for the N60, WIDE-S, WIDE-L, and N160 bands, respectively, where $F_{\rm Jy}$ denotes the expected model PSF flux in Jy (note that the fitting parameters are dependent on the flux units).
Following the adopted power-law FIS response function, the uncorrected-to-expected PSF flux ratio can be understood as
$F_{\rm uncorr}/F_{\rm PSF}=(\sum S_{ij,{\rm FIS}})/(\sum S_{ij}) = (\sum c S_{ij}^n)/(\sum S_{ij}) = \sum (c S_{ij}^n/S_{ij}) = \sum c S_{ij}^{(n-1)} \approx c F_{\rm PSF}^{(n-1)}$,
where $\sum$ stands for summation within some photometric contour.
Therefore, the fitting results above indicate that the archived FIS slow-scan maps have already been calibrated reasonably well indeed ($n \approx 0.9$ and $c \approx 1$). However, there still remain some residual effects that have not been completely mitigated.

The fact that the power-law index turned out to be less than unity ($n \approx 0.9$) suggests that the residual FIS response would cause overestimation for dimmer sources and underestimation for brighter sources. 
The crossover, at which the fit gray curve intersects with the horizontal dashed line of the ratio $=1$ in Figure\,\ref{rawphot}, occurs around $F_{\rm PSF} = (1/c)^{1/(n-1)} = 1.5, 7.0$, and 23\,Jy for the N60, WIDE-S, and WIDE-L bands, respectively (apparently no sources got overestimated in the N160 band).  
Thus, this observed residual detector response is consistent with the expectation that the slow-transient response of the Ge:Ga detector would cause underestimation for brighter objects.
However, we do not have enough information to identify the sources of the residual power-law FIS response, which is beyond the scope of the present study.
In Figure\,\ref{rawphot}, we also see that the uncorrected-to-expected flux ratio becomes greater than unity for dimmer PSFs.
This is somewhat counter-intuitive, but the observed residual power-law FIS response with the index $n \approx 0.9$ would cause overestimation, resulting in greater-than-nominal uncorrected fluxes after scaled up to recover fluxes within the 1-$\sigma$ contour of the super-PSF map.

When compared with the previous analysis done by \citet{Shirahata_2009}, the coefficients of the observed power-law response are much closer to unity and the indices of the power-law response are greater for the present FAST-processed data set.
This difference appears to stem from the difference of the map-making algorithm in SS-Tool and FAST. 
The final pixel values of the resulting map are determined by statistically analyzing all TSD readings occurring around the corresponding pixel positions.
With SS-Tool, equal weights are given to TSD readings within the beam size region around the pixel position (i.e., the PSF kernel is set to be a 2-D pillbox function with the beam-size top), while with FAST Gaussian weights are given to TSD readings that fall under the footprint of a Gaussian profile for which the FWHM equals to the beam size (set by  the Gaussian gridding convolution function mentioned in \S\,\ref{datareduction}).
This practically means that PSFs processed with SS-Tool tend to have a flatter, more suppressed peak than those processed with FAST. Because of the adopted 2-D pillbox PSF kernel profile, the measured PSF fluxes in the SS-Tool processed maps are reduced by as much as a few tens of \%.
We confirmed the effects of the PSF kernel shape to the resulting surface brightnesses by emulating the 2-D pillbox PSF profile with FAST.
For the present FAST processing, however, we adopted map-making parameters that are generally consistent with those used in processing FIS all-sky maps \citep{Doi_2015,Takita_2015}.
Hence, the present results indicate that FAST processing with the adopted parameters recovered considerably more fluxes than SS-Tool processing presented by \citet{Shirahata_2009}.
Nevertheless, the power-law response behavior of the detector still remains in the data and flux measurements suffer as much as about 50\% uncertainties.
Thus, we hereby confirm the need for surface brightness/flux correction even for the FAST-processed FIS data.

\subsection{\label{derive}Map Re-Scaling by the Inverse FIS Response Function}

In this section, we outline how the $(n,c)$ scaling parameter pair for the adopted FIS response function, $S_{ij, {\rm FIS}} = cS_{ij}^{n}$, are determined.\footnote{%
In the present derivation, it is assumed that surface brightnesses in FIS maps are given in the units of MJy\,sr$^{-1}$ (the default units of the {\sl AKARI\/}/FIS slow-scan maps) and derived fluxes are expressed in Jy.  
The derived $(n,c)$ parameters are valid only when the surface brightness units of the input {\sl AKARI\/} slow-scan maps ($S_{ij}$) are given in MJy\,sr$^{-1}$.}
The power-law exponent $n$ is equivalent to the slope of the logarithm of the FIS response function in the $\log(S_{ij, {\rm FIS}})$--$\log(S_{ij})$ space.
However, because we do not know $S_{ij}$ {\sl a priori}, we cannot simply determine $n$ by finding the slope of a line in the $\log(S_{ij, {\rm FIS}})$ vs.\ $\log(S_{ij})$ plot: we need a proxy for $S_{ij}$.
For the PSF/photometric references, we already know the expected fluxes, $F_{\rm PSF}$ = $\sum S_{ij}^{\rm PSF}$, from model predictions (Table\,\ref{obs_log}).
Hence, we can substitute $F_{\rm PSF}$ as a single-valued proxy for $S_{ij}^{\rm PSF}$ as long as there is a corresponding single-valued proxy for $S_{ij, {\rm FIS}}^{\rm PSF}$.
Of course, the uncorrected PSF flux, $F_{\rm PSF,FIS} = \sum S_{ij, {\rm FIS}}^{\rm PSF}$ would be the corresponding proxy.
However, as we already saw in the previous section, $F_{\rm PSF,FIS}$ via the contour photometry method are susceptible to low S/N ratios in general.
We can alternatively use the peak intensity of the PSF in the FIS map, $I_{\rm peak, {\rm FIS}}$, as a single-valued proxy for $S_{ij, {\rm FIS}}^{\rm PSF}$.
This is because 
(1) $I_{\rm peak, {\rm FIS}}$ should have achieved a sufficiently high S/N ratio when detected, and
(2) $F_{\rm PSF,FIS} = \sum S_{ij, {\rm FIS}}^{\rm PSF} = I_{\rm peak, {\rm FIS}} \sum f^{\prime}_{ij}$ for FIS PSF maps, for which $f^{\prime}_{ij}$ is well-defined due to the PSF scale invariance verified above (\S\,\ref{invariance}; Figure\,\ref{psfmad}).
Thus, we can constrain $n$ as the slope of a linear fit in the $\log(I_{\rm peak, {\rm FIS}})$ vs.\ $\log(F_{\rm PSF})$ plot.

Figure \ref{fig:separatenvalue} shows the $\log(I_{\rm peak,FIS})$--$\log(F_{\rm PSF})$ plots for all four {\sl AKARI\/} bands. 
The derived $n$ values are listed in Table\,\ref{nandcvalues}.
We emphasize here that the derived $n$ values are valid only when $I_{\rm peak, {\rm FIS}}$ is expressed in MJy\,sr$^{-1}$ and $F_{\rm PSF})$ in Jy.
The $F_{\rm PSF}$ values were taken from model predictions (Table\,\ref{obs_log}), 
while the $I_{\rm peak, {\rm FIS}}$ values were determined by fitting each of the PSF core by a 2-D Gaussian function.
Fitting of $\log(I_{\rm peak,FIS})$--$\log(F_{\rm PSF})$ was done with the Levenberg-Marquardt least-square minimization method, while uncertainties in both $F_{\rm PSF}$ and $I_{\rm peak,FIS}$ were properly propagated.
The $n$ values obtained in the present analysis turned out slightly smaller than unity ($0.90 \le n \le 0.96$). 
This behavior of $n$ can be understood straightforwardly as a progressive underestimation of the peak surface brightnesses for brighter objects, which would result in correspondingly underestimated flux values -- an expected manifestation of the slow transient response of the Ge:Ga detector (Figure\,\ref{rawphot}).

~\\
{\bf [Place Figure\,\ref{fig:separatenvalue} here]}\\

We then determined the $c$ value for each band.
Ideally, the FIS maps would recover the true surface brightness distribution of the sky when the maps are properly re-scaled by the inverse FIS response function. 
Hence, the re-scaled FIS maps would yield the expected fluxes of the PSF/photometric references via contour photometry, i.e., $F_{\rm PSF} \ge \sum S_{ij}^{\rm PSF} = \sum [(S^{\rm PSF}_{ij, {\rm FIS}}/c)^{1/n}] = c^{-1/n} \sum [(S^{\rm PSF}_{ij, {\rm FIS}})^{1/n}]$.
Here, it is worth noting that 
(1) the inequality symbol in the last equation refers to the fact that the sum is done only within the user-defined photometric contour as opposed to the idealized infinite aperture, and hence, the sum is always equal to or smaller than the expected/model PSF flux, $F_{\rm PSF}$, and
(2) this sum within the contour has to be done {\sl after} re-scaling the FIS map, because $\sum [(S^{\rm PSF}_{ij, {\rm FIS}})^{1/n}] \ne [\sum (S^{\rm PSF}_{ij, {\rm FIS}})]^{1/n}$ in general.
The $c$ factor is assumed to be unique for each band, and hence, can be pulled out of the sum. 
In practice, we fit the linear relation, $\sum_{3\sigma} [(S^{\rm PSF}_{ij, {\rm FIS}})^{1/n}] = c^{1/n} F_{\rm PSF,3\sigma}$, to derive $c^{1/n}$ as the slope of the fitted line.
Here, $3\sigma$ indicates that the corresponding values are evaluated/scaled within the 3-$\sigma$ contour (of the individual PSF map).
Again, we emphasize that the derived $c$ values are valid only when $S^{\rm PSF}_{ij, {\rm FIS}}$ is expressed in MJy\,sr$^{-1}$ and $F_{\rm PSF})$ in Jy.
The same Levenberg-Marquardt least-square minimization method was used in the fitting, considering uncertainties in both $\sum [(S^{\rm PSF}_{ij, {\rm FIS}})^{1/n}]$ and $F_{\rm PSF}$ that were properly propagated.
The $\sum [(S^{\rm PSF}_{ij, {\rm FIS}})^{1/n}]$--$F_{\rm PSF}$ plots for all four {\sl AKARI\/} bands are displayed in Figure \ref{fig:separatecvalue}.
The derived $c$ values are listed in Table\,\ref{nandcvalues}.

~\\
{\bf [Place Figure\,\ref{fig:separatecvalue} here]}\\
{\bf [Place Table\,\ref{nandcvalues} here]}\\

\subsection{Photometric Accuracy for PSF/Photometric Reference Sources}

With the $(n, c)$ parameter pair for each FIS band derived, we can now recover the corrected surface brightness distribution of the far-IR sky, $S_{ij}$, via the inverse FIS response function (eq.\ref{rescaling}).
Then, the corrected PSF fluxes are obtained as the sum of surface brightnesses within the adopted 3-$\sigma$ contour of the original PSF map, i.e., $F_{\rm PSF,3\sigma} = \sum_{3\sigma} [(S^{\rm PSF}_{ij, {\rm FIS}}/c)^{1/n}]$.
Usually, these fluxes would qualify perfectly valid ``measured" PSF fluxes.
Here, however, we would like to compare the corrected fluxes with the model fluxes.
Hence, the measured 3-$\sigma$ fluxes need to be scaled to represent fluxes in an idealized infinite aperture using the ratio between spatially-integrated normalized super-PSF profiles within the 3-$\sigma$ contour (of the original PSF map) and the 1-$\sigma$ contour (of the super-PSF map).
The quoted PSF flux values in Table\,\ref{FASTscaledphoto} are therefore ``measured" fluxes corresponding to the model fluxes.
Under the present scheme, the propagated uncertainties are combinations of uncertainties of the peak intensity fitting of the PSF profile (i.e., the MAD fluctuation), and of the background emission noise.
The uncertainties tend to be larger for dimmer objects in all bands.
This trending is most likely a manifestation of the Eddington bias (\cite{eddington_1913}; also more recently in context of sub-mm observations, \cite{Valiante_2016}), by which there are more fainter surface brightnesses to be scattered to brighter surface brightnesses by errors than brighter surface brightnesses to be scattered to fainter surface brightnesses by errors, yielding relatively greater errors for fainter surface brightnesses.

The corrected-to-expected PSF flux ratios as a function of the expected PSF flux for each FIS band are plotted in Figure\,\ref{fig:CalibrationRatio}.
Direct comparison between Figure\,\ref{rawphot} and Figure\,\ref{fig:CalibrationRatio} clearly indicates that the flux-dependent responsivity of the detector is now suppressed sufficiently.
The power-law fits of the corrected-to-expected PSF flux ratios are
$(1.08 \pm 0.04) \times F_{\rm Jy}^{(-0.02 \pm 0.01)}$,
$(1.07 \pm 0.02) \times F_{\rm Jy}^{(-0.03 \pm 0.01)}$,
$(0.98 \pm 0.03) \times F_{\rm Jy}^{(-0.03 \pm 0.01)}$, and
$(1.06 \pm 0.04) \times F_{\rm Jy}^{(-0.02 \pm 0.01)}$, respectively, for the N60, WIDE-S, WIDE-L, and N160 bands, where $F_{\rm Jy}$ is the 3-$\sigma$ expected flux in Jy (note that the fitting parameters are dependent on the flux units).
Compared with the corrected-to-expected PSF flux ratios (Figure\,\ref{rawphot}), these fits confirm that the ratios are very much improved to be closer to unity ($n \approx 0.98$ and $c \approx 1.05$) over a typical range of source fluxes greater than 0.1\,Jy (for SW) and 1\,Jy (for LW).
The flux accuracy of the PSF/photometric sources in terms of errors by the MAD with respect to their expected fluxes are 
$6 \pm 10$, 
$8 \pm 9$, 
$12 \pm 7$, and
$8 \pm 20$\%, 
respectively, for the N60, WIDE-S, WIDE-L, and N160 bands.


~\\
{\bf [Place Figure\,\ref{fig:CalibrationRatio} here]}\\
{\bf [Place Table\,\ref{FASTscaledphoto} here]}\\

\subsection{Photometric Accuracy for Compact Extended Sources}

\subsubsection{\label{extendedsources}%
Compact Extended Sources}

To ascertain the validity of the present surface brightness/flux correction method, we need to look at fluxes of compact extended sources.
For this purpose, we used {\sl AKARI\/} pointed observation slow-scan mapping data from the MLHES mission programme (excavating Mass Loss History in Extended dust shells of Evolved Stars; PI: I. Yamamura) and a few other smaller mission programmes aimed at mapping circumstellar shells\footnote{Other pointed observation mission programmes, FISPN (PI: P.\ Garc\'{\i}a-Lario), WRENV (PI: A.\ Marston), and OBSTR (PI: R.\ Blomme), were included.}.
This sub-set of {\sl AKARI\/} slow-scan maps consists of 157 compact extended sources of various evolutionary status from early to late types in the present analysis and constitutes the most complete {\sl AKARI\/}/FIS far-IR slow-scan imaging data set of the circumstellar shells with a variety of target types and fluxes.
All of slow-scan observations for these compact extended sources were carried out with the AOT FIS01 scan sequence, and their TSD data were processed in the same way as the PSF/photometric standards.
Then, the resulting FIS maps were re-scaled using the inverse FIS response function with the derived $(n, c)$ parameters and corrected surface brightnesses were summed up within 3-$\sigma$ contours as established above.
To perform a direct comparison with published monochromatic flux densities derived from other projects, we also need to color-correct the derived flux values of these compact extended sources following the procedure elucidated by \citet{Shirahata_2009}, assuming a 5,000\,K blackbody for simplicity (i.e., dividing the derived flux by 
$1.05 \pm 0.01$,
$1.40 \pm 0.01$,
$0.937 \pm 0.001$, and
$0.993 \pm 0.001$, 
for the N60, WIDE-S, WIDE-L, and N160 band, respectively.
Note that the correction factors do not vary more than 0.5\,\% from 1,000 to 10,000\,K.)

\subsubsection{%
Comparison with {\sl IRAS\/} Fluxes}

Published fluxes of these compact extended sources were looked up from the Infrared Astronomical Satellite ({\sl IRAS\/}) Point Source Catalog (PSC), version 2.0 \citep{iras}.
We found 140 and 129 entries in the {\sl IRAS\/} 60 and 100\,$\micron$ bands, corresponding to the {\sl AKARI\/} N60 and WIDE-S bands, respectively.
These {\sl IRAS\/} fluxes were also color-corrected assuming the same 5,000\,K blackbody.
Additional correction to take into account for the slightly offset bandpasses between the two instruments (i.e., scaling by the flux ratio of the assumed 5,000\,K blackbody at the central wavelengths of the instruments) was applied \citep{muller_2011}.
The top two frames of Figure\,\ref{fig:fluxcomparison} show the correlation between color-corrected {\sl IRAS\/} and {\sl AKARI\/}/FIS fluxes ({\sl IRAS\/} 60\,$\micron$ vs.\ N60 and {\sl IRAS\/} 100\,$\micron$ vs.\ WIDE-S).
These correlations can be fit into the following power-law relations:
$F_{\rm IRAS60} = (1.60 \pm 0.02) F_{\rm N60}^{(0.87 \pm 0.01)}$ and
$F_{\rm IRAS100} = (2.45 \pm 0.02) F_{\rm WIDE-S}^{(0.80 \pm 0.01)}$ (where $F$s denote the flux in the corresponding {\sl AKARI\/} and {\sl IRAS\/} band in Jy).
Note that the detection limit in the {\sl IRAS\/} 100\,$\micron$ band is about 1\,Jy and in the plot many sources dimmer than the detection limit were identified as having the limiting flux due to the Eddington bias.

The observed general correspondence between {\sl IRAS\/} and {\sl AKARI\/} (13 and 14\,\% rms error in the N60 and WIDE-S bands, respectively) demonstrates that the present surface brightness correction method for {\sl AKARI\/} slow-scan maps yields reasonable results, while closer inspection suggests that {\sl IRAS\/} fluxes tend to be overestimated more for dimmer sources and underestimated more for brighter sources.
We attribute this trending to the spatial resolution of and the way fluxes are measured by {\sl IRAS\/}.
We remind readers, however, that the present comparison is made between {\sl AKARI\/} fluxes of intrinsically extended objects and their corresponding flux entries in the {\sl IRAS\/} PSC.
The nominally large $6^{\prime}$ resolution of {\sl IRAS} \citep{iras} would yield fluxes that include the circumstellar component for brighter sources (hence, relatively good correlation), but for very bright sources their strong nebula flux component tends to be underestimated (hence, lower {\sl IRAS\/} fluxes).
Meanwhile, {\sl IRAS\/} PSC fluxes are measured by comparing the time-series signal profile of the {\sl IRAS\/} detector with point-source signal profile templates \citep{iras}.
This is practically equivalent to aperture photometry, and hence, it is possible to have flux overestimates for dimmer sources (hence, greater {\sl IRAS\/} fluxes) by the same reason discussed earlier.

\subsubsection{%
Comparison with {\sl Herschel\/}/PACS Fluxes}

We also obtained flux measurements made with the Photodetector Array Camera and Spectrometer (PACS; \cite{Poglitsch_2010}) on-board the Herschel Space Observatory ({\sl Herschel\/}; \cite{Pilbratt_2010}). 
Using the Herschel Science Archive\footnote{http://www.cosmos.esa.int/web/herschel/science-archive}, we identified compact extended sources mapped by both {\sl AKARI\/} and {\sl Herschel\/}, many of which were observed as part of the MESS key project \citep{mess}.
The PACS instrument has three broadbands at 70, 100, and 160\,$\micron$ that were determined to have been calibrated to yield the photometric accuracy of 5\% error at best \citep{Ali_2011,Sauvage_2011,Paladini_2012,balog_2014}: hence, PACS fluxes for these compact extended sources would provide an independent baseline for comparison with the present {\sl AKARI\/}/FIS fluxes.

We used the Herschel Interactive Processing Environment (HIPE, version 13; \cite{hipe}) and Scanamorphos data reduction tool (Scanamorphos, version 21; \cite{scana}) to generate {\sl Herschel} broadband images of these common compact extended sources.
First, the archival data were processed with HIPE from level 0 to level 1 (basic pipeline reduction steps including corrections for instrumental effects), and then, the level 1 data were ingested into Scanamorphos, with which corrections for electronic instabilities, deglitching, flux calibration, and map projection were performed. 
Scanamorphos was chosen because it would reconstruct surface brightness maps of extended sources with the lowest noise. 
After {\sl Herschel\/}/PACS images at 70 and 160\,$\micron$ were obtained, we defined the photometric contour at the 3-$\sigma$ level and summed up the pixel counts (Jy\,pixel$^{-1}$) within the contour into flux (Jy).
These steps of data reduction and photometry follow those described by \citet{Ueta2014}.

The derived uncorrected {\sl Herschel\/}/PACS fluxes assume a flat spectrum (i.e., $\nu F_{\nu} = \lambda F_{\lambda} = \textrm{const.}$) at the center wavelength of each band. 
This allows fluxes measured with one instrument to be interpolated to those measured by the other instrument for comparisons after an appropriate color correction, which is needed to reflect the fact that the actual source spectrum is not flat.
To keep the consistency, we assumed the same 5,000\,K blackbody for simplicity and adopted the color correction factors and the additional correction factor to translate monochromatic flux at the PACS 70, 100, and 160\,$\micron$ band to those at the N60, WIDE-S, and WIDE-L bands, respectively \citep{muller_2011}.

The bottom four frames of Figure\,\ref{fig:fluxcomparison} shows comparison between the appropriately-corrected {\sl AKARI\/}/FIS and {\sl Herschel\/}/PACS fluxes.
These correlations can be fit with
$F_{\rm PACS70}   = (1.11 \pm 0.01) F_{\rm N60}^{(1.01 \pm 0.01)}$,
$F_{\rm PACS100} = (1.39 \pm 0.01) F_{\rm WIDE-S}^{(0.92 \pm 0.01)}$,
$F_{\rm PACS160} = (1.45 \pm 0.01) F_{\rm WIDE-L}^{(1.01 \pm 0.01)}$, and
$F_{\rm PACS160} = (1.48 \pm 0.01) F_{\rm N160}^{(0.89 \pm 0.01)}$, 
where $F$s denote the flux in the corresponding {\sl AKARI\/} and {\sl Herschel\/}/PACS bands in Jy.
Given the crude assumption in the comparison (5,000\,K blackbody color-correction), 
the observed correlations exhibit consistency between {\sl AKARI\/}/FIS and {\sl Herschel\/}/PACS fluxes (13, 29, 24, and 20\,\% rms error in the N60, WIDE-S, WIDE-L, and N160 bands, respectively) and suggest that {\sl AKARI\/}/FIS and {\sl Herschel\/}/PACS measurements are mutually reproducible in general.
Thus, we conclude that the present surface brightness correction method for the {\sl AKARI\/}/FIS slow-scan maps works properly to allow investigations into the surface brightness distributions of compact extended sources.

~\\
{\bf [Place Figure\,\ref{fig:fluxcomparison} here]}\\

\subsection{Flux Scaling in the Point-Source and Compact-Extended-Source Corrections}

\citet{Shirahata_2009} presented a flux correction method for point sources based on the ``total flux," which is the flux of the target source {\sl plus} contributions from the sky and dark current (i.e., everything that is detected by the FIS arrays; eq.\,1 of \cite{Shirahata_2009}).
This is the only flux correction method for {\sl AKARI\/}/FIS slow-scan maps verified by the {\sl AKARI\/} team, but the method was verified only for point sources.
Because of lack of an appropriate flux correction method, photometric measurements previously performed for compact extended sources detected with {\sl AKARI\/}/FIS were corrected for (or not corrected for) on a case-by-case basis (e.g.\ \cite{Kaneda_2007,Hirashita_2008,Ueta_2008,Kaneda_2010,Cox_2011,Izumiura_2011}).
Now, it is possible to assess how the quality of photometric measurements may be compromised by indiscriminately adopting this point-source flux correction method for sources that are extended.

To do this assessment, we constructed FAST-processed maps of the MLHES target sources {\sl without} dark-subtraction and median-filtering (and {\sl without} re-scaling with the FIS response function) to emulate the ``total flux" and determine flux correction scaling factors accordingly \citep{Shirahata_2009}.
These flux correction factors were then applied to uncorrected flux measurements made from FAST-processed maps {\sl with} dark-subtraction and median-filtering (but {\sl without} re-scaling with the FIS response function) via {\sl contour photometry} within the 3-$\sigma$ contour, which defines the ``aperture" for this set of measurements.
Here, readers are reminded again that for the MLHES target sources that are potentially intrinsically extended, aperture correction cannot be performed.
For lack of a better descriptor, however, we would keep referring these fluxes as ``aperture photometry" fluxes to distinguish them from ``contour photometry" fluxes based on the present method.
We also measured ``uncorrected" fluxes by adopting the same 3-$\sigma$ contours with the uncorrected slow-scan maps (i.e., archived data processed with FAST {\sl without} re-scaling with the FIS response function).

We then compared these three sets of fluxes by making two sets of ratios, the ratios of the aperture-photometry and uncorrected fluxes to the contour-photometry fluxes.
In Figure\,\ref{fig:ShirahataCompare}, these ``aperture-to-contour" and ``uncorrected-to-contour" photometry ratios (dots and crosses, respectively) are plotted as a function of the contour-photometry flux.
The loci of points exhibit the power-law variation that is not removed by the application of the map re-scaling with the FIS response function.
The median photometry ratios are found to be
$1.49 \pm 0.07$,
$1.77 \pm 0.12$, 
$2.26 \pm 0.19$, and
$1.74 \pm 0.06$
for the aperture- to contour-photometry comparison, and 
$0.83 \pm 0.09$,
$1.03 \pm 0.12$
$1.26 \pm 0.11$, and
$0.48 \pm 0.02$
for the uncorrected to contour-photometry comparison,
for the N60, WIDE-S, WIDE-L, and N160 bands, respectively.
Hence, this assessment confirms that the aperture-photometry fluxes always overestimate at least 50\% up to more than double and that the uncorrected fluxes underestimate (20\% to 50\%) or there remains a strong brightness dependence (up to 40\% over/underestomates) even when the uncorrected fluxes are in the right ballpark for compact extended sources.
Thus, it is demonstrated that measured fluxes of compact extended sources can suffer greatly from 150 to 230\% overestimates or 20 to 50\% underestimates if the present scheme of surface brightness correction for the archived {\sl AKARI\/}/FIS slow-scan maps is not followed.

~\\
{\bf [Place Figure\,\ref{fig:ShirahataCompare} here]}\\

\section{\label{summary}%
Conclusions}

\begin{figure}
\begin{center}
\includegraphics[width=\textwidth]{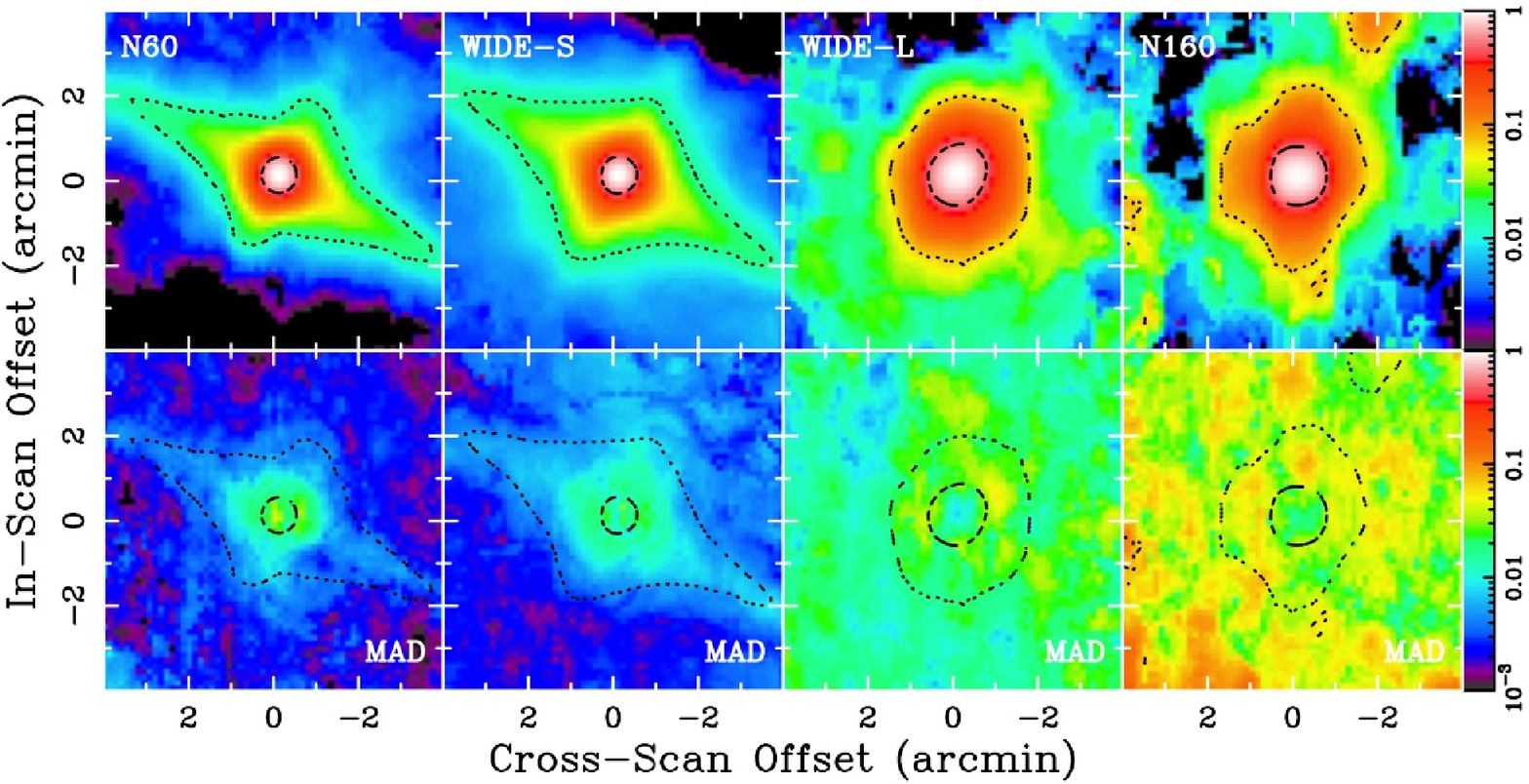}
\caption{\label{psfmad}%
The {\sl AKARI\/}/FIS super-PSF images (top row) and the corresponding median absolute deviation (MAD) maps (bottom row) in the N60, WIDE-S, WIDE-L, and N160 bands (top row; from left to right).
The super-PSF maps are made by taking the normalized median of the observed PSF reference maps.
The logarithmic color scaling of the images, from 0.1\,\% to 100\,\% relative to the peak intensity, is indicated in the wedge on the right.
The dashed and dotted contours in the PSF surface brightness distribution images represent the FWHM and 5\,$\sigma$ levels, respectively.}
\end{center}
\end{figure}

\begin{figure}
\begin{center}
\includegraphics[width=\textwidth]{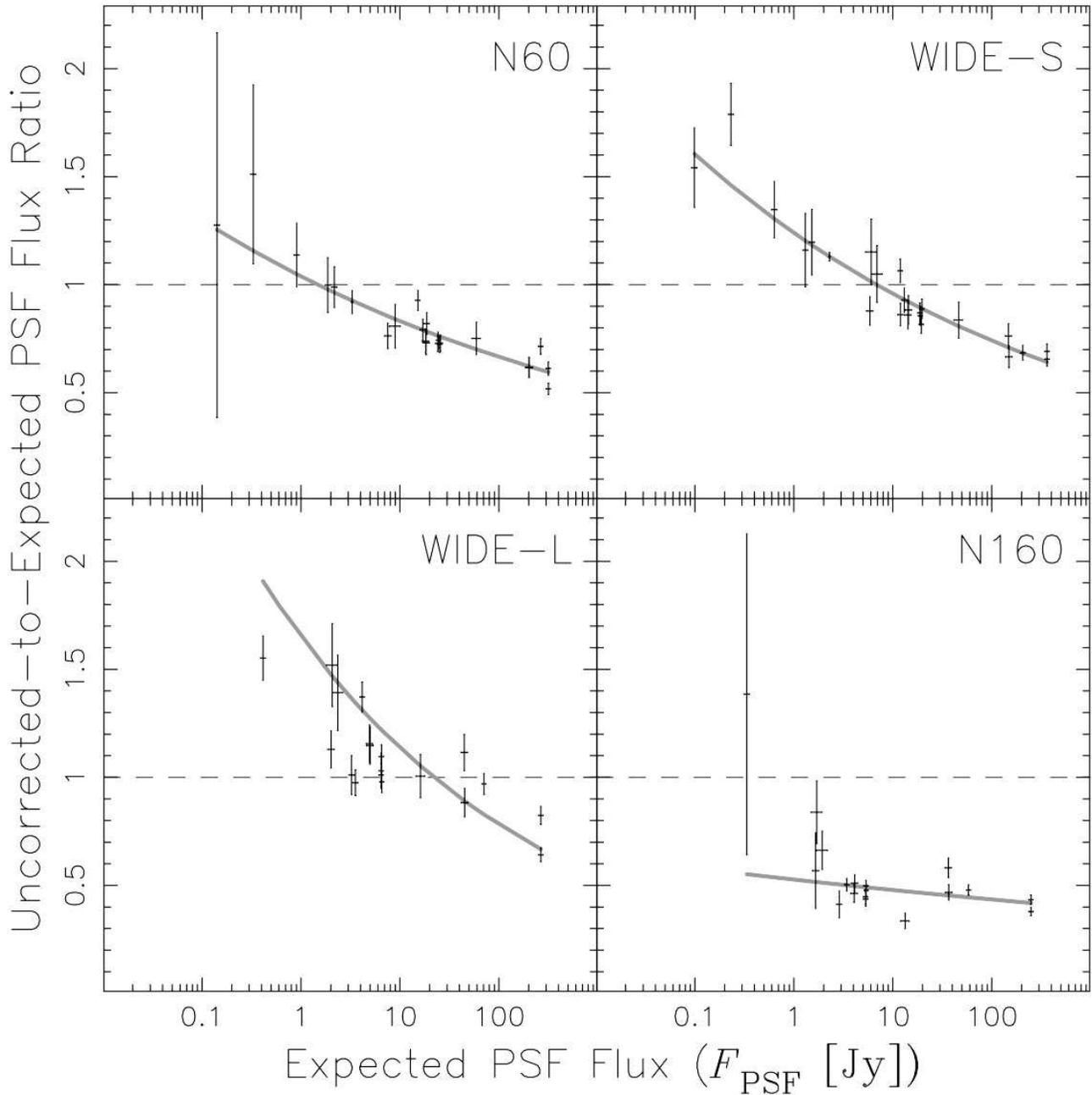}
\caption{\label{rawphot}%
The uncorrected-to-expected PSF flux ratio as a function of the expected PSF flux in each of the {\sl AKARI\/}/FIS bands.
The gray solid lines show the power-law best fits.
These plots confirm that the power-law response of the FIS detector with respect to the source flux is still present in the FAST-processed raw/archived FIS maps.
These fluxes are measured by summing all surface brightness pixel counts within the 3-$\sigma$ aperture.}
\end{center}
\end{figure}

\begin{figure}
\begin{center}
\includegraphics[width=\linewidth]{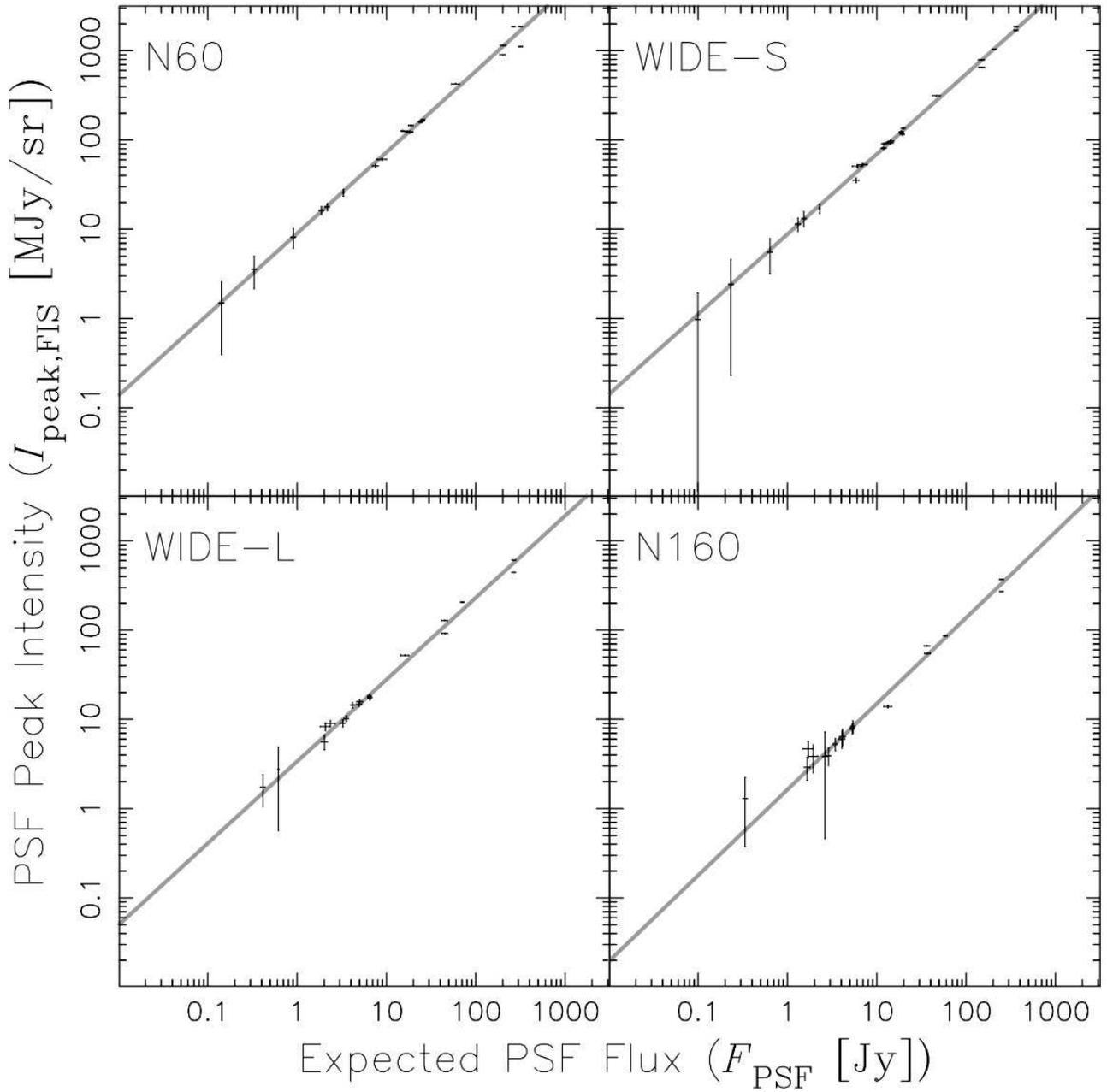}
\caption{\label{fig:separatenvalue}%
The correlation between the PSF peak intensity ($I_{\rm peak,FIS}$ in MJy/sr, as measured in the FAST-processed FIS maps) and the corresponding expected PSF flux ($F_{\rm PSF}$ in Jy; Table\,\ref{obs_log}).
The slope of the linear fit (shown as a gray line) defines the power-law index $n$ of the adopted FIS response function, $S_{ij, {\rm FIS}} = \mathcal{R}(S_{ij}) = c S_{ij}^n$. 
See Table \ref{nandcvalues} for the derived $n$ values.}
\end{center}
\end{figure}

\begin{figure}
\begin{center}
\includegraphics[width=\linewidth]{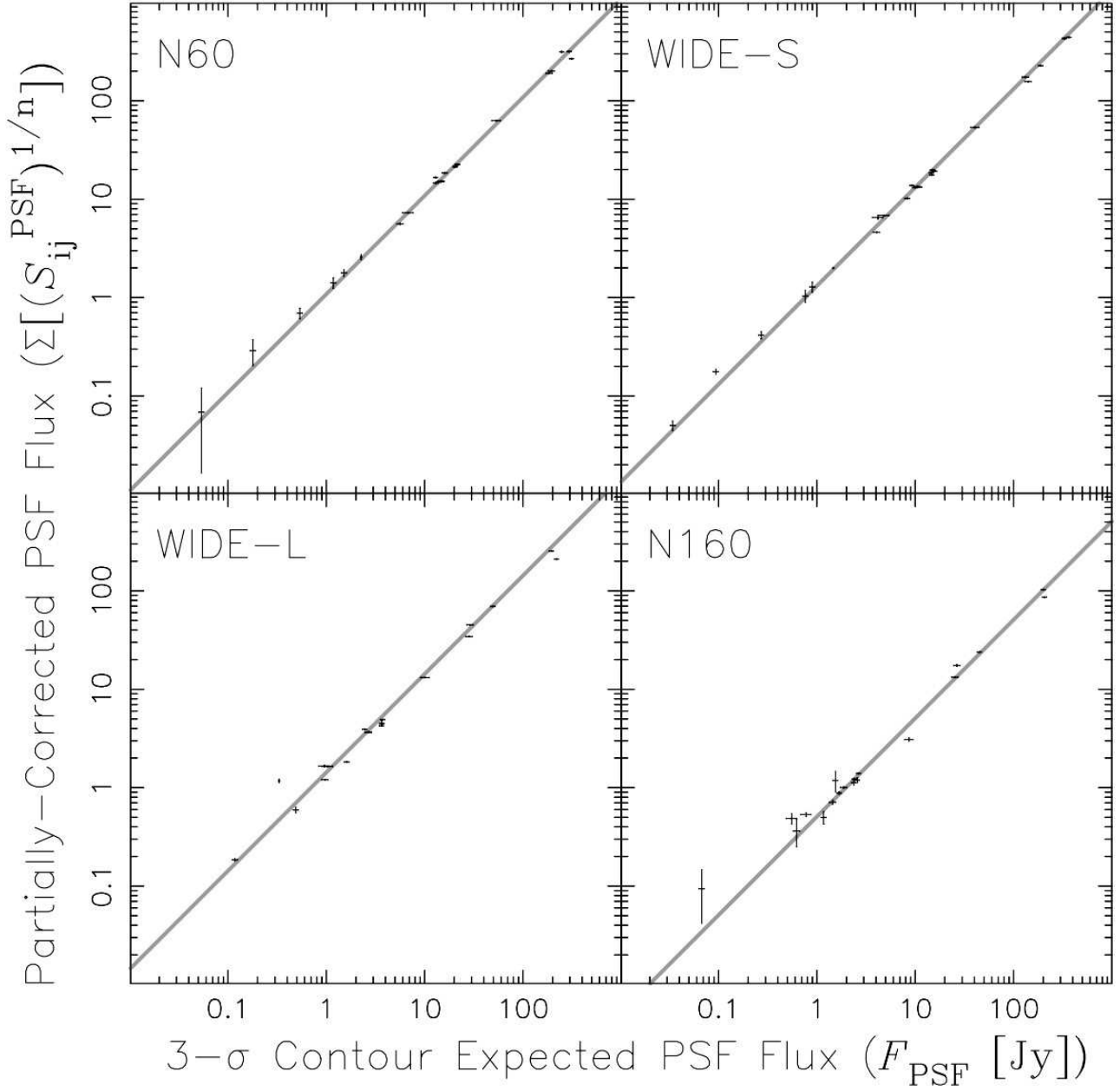}
\caption{\label{fig:separatecvalue}%
The correlation between the partially-corrected PSF flux 
($\sum_{3\sigma} [(S^{\rm PSF}_{ij, {\rm FIS}})^{1/n}]$) measured within the 3-$\sigma$ contour in the FAST-processed FIS maps re-scaled by the inverse FIS response function) and the corresponding expected flux scaled to the 3-$\sigma$ contour ($F_{\rm PSF},3\sigma$ in Jy; Table\,\ref{obs_log}).
The slope of the linear fit (shown as a gray line) determines $c^{1/n}$, from which the power-law coefficient $c$ of the adopted FIS response function, $S_{ij, {\rm FIS}} = \mathcal{R}(S_{ij}) = c S_{ij}^n$ is computed. 
See Table \ref{nandcvalues} for the derived $c$ values.}
\end{center}
\end{figure}

\begin{figure}
\begin{center}
\includegraphics[width=\linewidth]{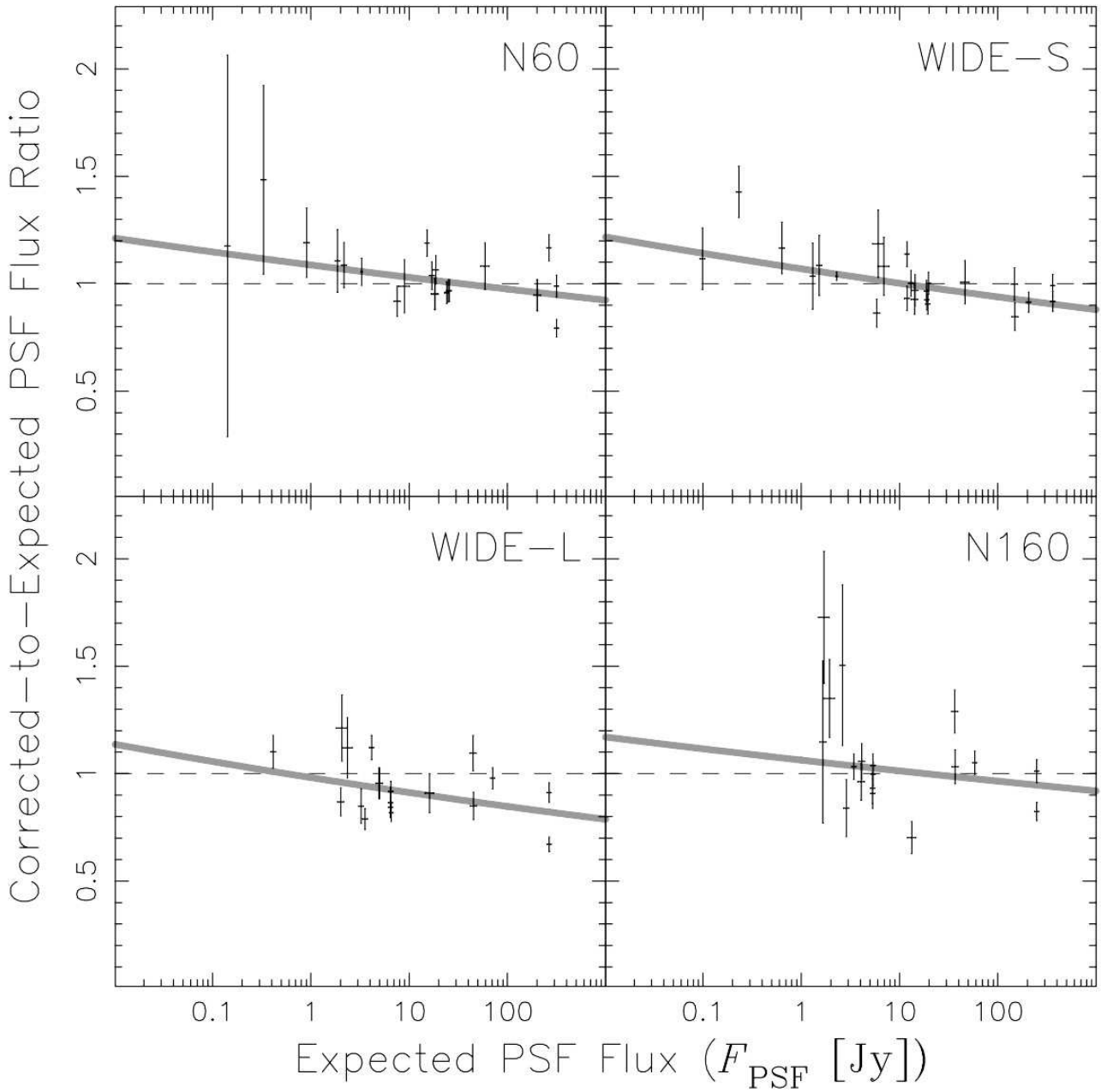}
\caption{\label{fig:CalibrationRatio}%
The corrected-to-expected PSF flux ratios as a function of the expected PSF flux in each of the {\sl AKARI\/}/FIS bands.
The gray solid lines show the power-law best fits.
These plots confirm that the surface brightness correction applied to the raw/archived FIS maps suppressed the signal-strength-dependent response of the FIS detector in the FAST-processed FIS maps.}
\end{center}
\end{figure}

\begin{figure}
\begin{center}
\includegraphics[width=0.8\linewidth]{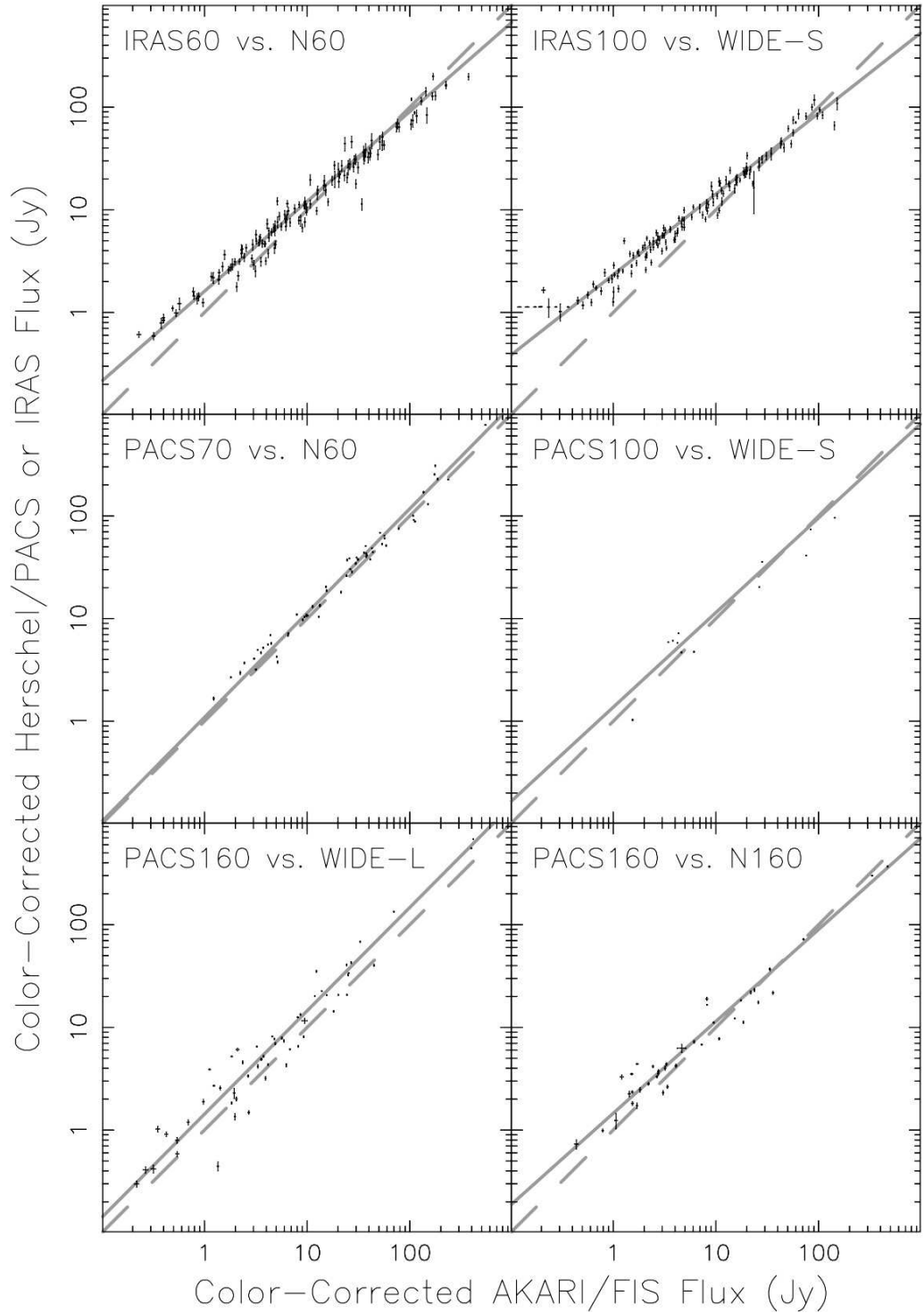}
\caption{\label{fig:fluxcomparison}%
Comparisons between {\sl AKARI\/}/FIS fluxes of compact extended sources and (1) {\sl IRAS} fluxes (top two frames; {\sl IRAS\/} 60\,$\micron$ vs.\ N60 and {\sl IRAS\/} 100\,$\micron$ vs.\ WIDE-S) and (2) {\sl Herschel\/}/PACS fluxes (bottom four frames; 
PACS\,70\,$\micron$ vs.\ N60, 
PACS\,100\,$\micron$ vs.\ WIDE-S, 
PACS\,160\,$\micron$ vs.\ WIDE-L, and
PACS\,160\,$\micron$ vs.\ N160).
These fluxes are color-corrected assuming a 5,000\,K blackbody for simplicity and also corrected for the bandpass offset. 
Solid gray lines are power-law fits of the correlation, while dashed lines are the exact match.}
\end{center}
\end{figure}

\begin{figure}
\begin{center}
\includegraphics[width=\linewidth]{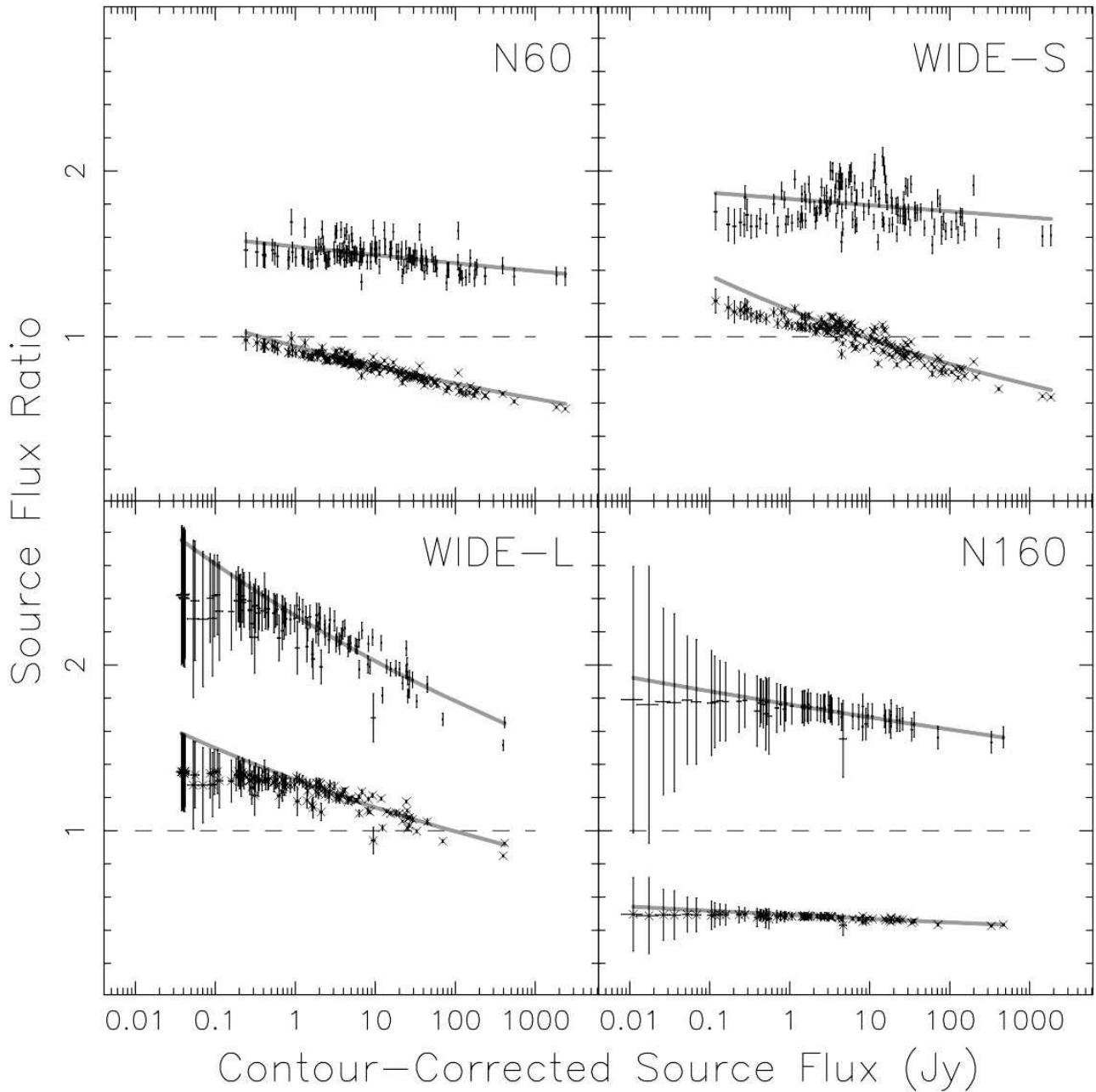}
\caption{\label{fig:ShirahataCompare}%
Comparisons of fluxes of compact extended sources obtained by the aperture photometry, contour photometry with the map re-scaling, and contour photometry without the map re-scaling expressed by the flux ratios.
The aperture- to contour-photometry (with the map re-scaling) ratios are dots with error bars (the top locus of points), while the uncorrected to contour-photometry corrected ratios are crosses with error bars (the bottom locus of points).
Gray lines are power-law fits.
}
\end{center}
\end{figure}

We established a general method to re-calibrate {\sl AKARI\/}/FIS slow-scan surface brightness images with the inverse FIS response function.
The purpose of this scheme is to recover the correct surface brightness distribution of compact extended sources, which are more extended than point sources but less extended than diffuse background, and consequently, to derive their fluxes as a simple sum of the surface brightnesses within the contour encircling the detected extent of the target sources.
This method is general and applicable to any source (including point sources) provided that the source is not considered diffuse (i.e., less extended than about 10$^{\prime}$, which is the nominal single scan angular width). 
Anticipating the public release of {\sl AKARI\/}/FIS slow-scan maps along with FAST expected to happen in around April 2017 (Takita et al.\ {\sl in preparation}), those who wish use {\sl AKARI\/}/FIS maps for their science that involves objects that are extended but more compact than about $10^{\prime}$ are encouraged to adopt this correction scheme to obtain correct surface brightness distributions of the target sources, especially when the data are taken in the same observing mode.
The results of the present investigations are summarized below:

\begin{enumerate}

\item{%
This method is based on the empirical power-law FIS response function, $\mathcal{R}$, of the form $S_{ij, {\rm FIS}} = \mathcal{R}(S_{ij}) = c S_{ij}^n$, which relates the archived raw/uncorrected FIS maps ($S_{ij, {\rm FIS}}$) and the true surface brightness distribution of the far-IR sky ($S_{ij}$).
In this formulation, the corrected surface brightness distribution, $S_{ij}$, is recovered by using the inverse power-law FIS response function, $\mathcal{R}^{-1}$, via $S_{ij} = \mathcal{R}^{-1}(S_{ij, {\rm FIS}}) = (S_{ij, {\rm FIS}}/c)^{1/n}$.
Then, the corrected flux of a target source is measured via contour photometry performed with the re-scaled map, i.e., $F_{\rm SRC} = \sum S_{ij} = \sum [(S_{ij, {\rm FIS}}/c)^{1/n}]$. 
This method is valid as long as the target source is not extended more than $\sim$ 10$^{\prime}$ (otherwise the source emission is considered to be equivalent to diffuse background, to which the archived {\sl AKARI\/}/FIS TSDs are already absolutely calibrated).}

\item{%
The PSF shape of the {\sl AKARI\/}/FIS slow-scan maps is fairly uniform irrespective of the brightness of the PSF/photometric reference source (the PSF scale invariance; Figure\,\ref{psfmad}).
This scale invariance of the {\sl AKARI\/}/FIS PSFs guarantees the power-law nature of the FIS response function, $\mathcal{R}$.}

\item{%
The FIS AKARI Slow-scan Tool (FAST) allows more flexible/extensive processing/correction of the {\sl AKARI\/}/FIS slow-scan TSDs (mainly concerned with de-glitching and cal-lamp after-effect correction) and produces slow-scan maps with the reduced amount of artifacts and noise without compromising the PSF shape stability and scale invariance.}

\item{%
The $(n, c)$ parameter pair for the adopted FIS response function, $\mathcal{R}$, are determined when surface brightnesses in FIS maps are given in MJy\,sr$^{-1}$ (and fluxes are expressed in Jy: Table\,\ref{nandcvalues}), based on photometric measurements of the PSF/photometric reference sources (Table\,\ref{obs_log}) done via contour photometry with the 3-$\sigma$ photometric contour.
Hence, to apply the present correction/re-scaling with the FIS response function and the $(n, c)$ parameters, the input {\sl AKARI\/}/FIS slow-scan maps must have the units of MJy\,sr$^{-1}$.
The overall photometric accuracy of the present correction method for compact extended sources is determined to be generally better than 10\% error.}

\item{%
The re-scaled FIS maps ($S_{ij}$) present the correct surface brightness distribution of the extended target sources.
Hence, their fluxes must be obtained via contour photometry as a simple sum of pixel surface brightness values within an adopted contour (i.e., $F_{\rm SRC} = \sum S_{ij}$).
If the flux correction method previously developed for point sources \citep{Shirahata_2009} is indiscriminately used for compact extended sources, one may overestimate fluxes by up to 230\%.
Meanwhile, if no re-scaling is performed for the archived FIS data , one may underestimate fluxes of compact extended sources by up to 50\%.
Without proper re-scaling of the archived FIS data, one will never recover the correct surface brightness distribution of the detected sources.
For this reason, {\sl AKARI\/}/FIS maps of compact extended sources should be corrected for the method presented here.}

\item{%
The particular correction parameters quoted here (Table\,\ref{nandcvalues}) are derived when the archived {\sl AKARI\/}/FIS TSDs are processed with FAST using the Gaussian gridding convolution function of the beam (FWHM) size of $30^{\prime\prime}$ and $50^{\prime\prime}$ for the SW and LW bands, respectively (with the {\sl searching\_radius} being $\times 2$ of the beam), the 5-$\sigma$ clipping (with the {\sl clipping\_radius} being the beam size), and median-filter with the 250\,s width
(see \cite{Verdugo_2007} and \cite{ikeda_2012a} for more details).
Because the PSF shape of the resulting FIS maps (i.e., $f^{\prime}_{ij}$) can vary depending on the choice of these map-making and PSF kernel parameters, 
the correction parameters ($n$, $c$) need to be re-examined by following the procedure presented here when the PSF profiles are significantly different from those presented here (Figure\,\ref{psfmad}).}

\end{enumerate}

\begin{ack}
This research is based on observations with AKARI, a JAXA project with the participation of ESA.
TU recognizes a partial support from the Japan Society for the Promotion of Science (JSPS) through a FY2013 long-term invitation fellowship program.
RLT was partially supported by the NSF EAPSI Program (OISE-1209948) and thank JAXA/ISAS and OAO for use of their facilities while conducting part of this research in Japan.
The authors thank the anonymous referee for helpful suggestions. 
\end{ack}

\clearpage

\clearpage

\clearpage

\clearpage

\clearpage

\clearpage

\clearpage
{\small\setlength\tabcolsep{3pt}
\begin{longtable}{lcccrrrrc}
\caption{ \label{obs_log}PSF/Photometric Reference Objects and Their Model Fluxes$^{*}$} 
\hline
\multicolumn{4}{l}{} &
\multicolumn{5}{c}{Model Flux$^{\dagger}$}\\
\cline{5-9}
\multicolumn{4}{c}{Observations} &
\multicolumn{1}{c}{N60} & 
\multicolumn{1}{c}{WIDE-S} & 
\multicolumn{1}{c}{WIDE-L} & 
\multicolumn{1}{c}{N160} & 
\multicolumn{1}{c}{Accuracy}\\
\cline{1-4}  
\multicolumn{1}{c}{Target} & 
\multicolumn{1}{c}{OBSID} & 
\multicolumn{1}{c}{Date} & 
\multicolumn{1}{c}{AOT Params$^{\ddagger}$} & 
\multicolumn{1}{c}{[Jy]} & 
\multicolumn{1}{c}{[Jy]} & 
\multicolumn{1}{c}{[Jy]} & 
\multicolumn{1}{c}{[Jy]} & 
\multicolumn{1}{c}{[\%]}\\
\hline
\endfirsthead
\multicolumn{9}{l}{\scriptsize $^{*}$ This table is a subset of Table 2 of \citet{Shirahata_2009}.}\\
\multicolumn{9}{l}{\scriptsize $^{\dagger}$ Model calculation: Stars: Cohen et al. (1999, 2003a, 2003b); Asteroids \& Planets: Muller and Lagerros (1998, 2002); Moreno (1998).}\\
\multicolumn{9}{l}{\scriptsize $^{\ddagger}$ AOT FIS01 parameters: detector reset interval (s); scan speed ($^{\prime\prime}$\,s$^{-1}$); scan spacing ($^{\prime\prime}$).}\\
\endlastfoot
\multicolumn{9}{c}{Stars}\\
\hline
$\alpha$ CMa & 5110034\_001 & 2006/10/07 18:28:06 & 2.0;8;70 & 3.290 &  2.293& 0.616 & 0.497 & 1.5\\
$\alpha$ Boo & 5110039\_001 & 2007/01/15 00:02:26 & 1.0;8;70 & 18.689&  13.089& 3.558 & 2.879 & 6\phantom{.5}\\
$\alpha$ Tau &5110045\_001 & 2007/02/28 14:18:57 & 1.0;8;70 & 17.042&  11.939& 3.249 & 2.630 & 6\phantom{.5}\\
HD 216386 &5110068\_001 &  2007/06/03 01:05:46 & 2.0;8;70 & 2.177 & 1.524 & 0.414 & 0.335 & 6\phantom{.5}\\
HD 98118 & 5110072\_001& 2007/06/10 01:17:14 & 2.0;8;70 & 0.330 & 0.323 & 0.063 & 0.051 & 6\phantom{.5}\\
HD 222643 & 5110075\_001 & 2007/06/11 01:23:08 & 2.0;8;70 & 0.142 & 0.099 & 0.027 & 0.022 & 6\phantom{.5}\\
HD 224935 & 5110070\_001 & 2007/06/20 00:48:29 & 2.0;8;70 & 1.869 & 1.309 & 0.355 & 0.288 & 6\phantom{.5}\\
HD 92305 & 5110092\_001 & 2007/08/23 12:12:43 & 2.0;8;70 & 0.906 & 0.636 & 0.173 & 0.140 & 6\phantom{.5}\\
\hline
\multicolumn{9}{c}{Asteroids and Planets}\\
\hline
241 Germania & 5011065\_001 & 2006/04/27 15:44:31 & 0.5;8;70 & 8.958 &  6.932& 2.356 & 1.940 & 12.5\\
\multicolumn{1}{c}{\dots} &  5011165\_001  & 2006/04/27 23:59:07 & \dots & 7.813 &  6.064& 2.073 & 1.707 & 12.5\\
6 Hebe &5011066\_001 &  2006/04/30 03:07:09 & 0.5;8;70 & 25.258&  19.382& 6.469 & 5.313 & 5\phantom{.5}\\
\multicolumn{1}{c}{\dots} & 5011166\_001        & 2006/05/01 00:34:26 & \dots & 25.681&  19.699& 6.570 & 5.396 & 5\phantom{.5}\\
511 Davida & 5011067\_001 & 2006/05/02 22:50:20 &  0.5;8;70 & 18.394&  14.387& 4.999 & 4.127 & 7.5\\
\multicolumn{1}{c}{\dots} & 5011167\_001 &         2006/05/03 12:02:32 & \dots & 18.185&  14.214& 4.933 & 4.071 & 7.5\\
2 Pallas & 5110027\_001 & 2006/09/27 06:20:31 & 0.5;8;70 & 59.254&  46.375& 16.142 & 13.329 & 10\phantom{.5}\\
1 Ceres &5110032\_001 &  2006/11/08 14:58:11 & 0.5;8;70 & 264.848& 206.126&  70.786&  58.327& 5\phantom{.5}\\
93 Minerva & 5110033\_001 & 2006/11/20 00:42:13 & 1.0;8;70 & 7.551 &  5.873& 2.107 & 1.662 & 7.5\\
65 Cybele & 5110038\_001 & 2006/12/28 00:16:17 & 1.0;8;70 &  15.192&  11.905& 4.155 & 3.431 & 5\phantom{.5}\\
4 Vesta & 5110047\_001 & 2007/02/23 22:33:11 & 0.5;8;70 & 200.598 & 147.871 &  44.748& 36.486 & 7.5\\
\multicolumn{1}{c}{\dots} & 5110046\_001 & 2007/02/24 00:12:31 & 0.5;15;70 &   202.519& 149.228&  45.113& 36.778 & 7.5\\
52 Europa & 5110058\_001 & 2007/04/14 23:08:31 & 0.5;8;70 & 24.150&  18.807& 6.467 & 5.328 & 5\phantom{.5}\\
\multicolumn{1}{c}{\dots} & 5110059\_001 & 2007/04/15 22:19:51 & 0.5;15;70 & 24.328&  18.941& 6.511 & 5.364 & 5\phantom{.5}\\
Neptune &5110066\_001 &  2007/05/13 01:22:57  & 0.5;8;70 & 315.942& 361.867& 265.605& 248.897& 5\phantom{.5}\\
\multicolumn{1}{c}{\dots} & 5110067\_001 &  2007/05/13 19:36:26 & 0.5;15;70 & 316.215& 362.171& 265.833& 249.113& 5\phantom{.5}\\
\hline
\end{longtable}}

\clearpage

{
\begin{longtable}{lcccr}
\caption{ \label{nandcvalues}Parameters and Characteristics of Flux Correction for Contour Photometry} \\
\hline
\multicolumn{1}{l}{} & \multicolumn{2}{c}{Power-Law Fit Parameters} & \multicolumn{1}{c}{Correction Accuracy} & \multicolumn{1}{c}{Flux Range} \\
\cline{2-3}
\multicolumn{1}{l}{Band} & \multicolumn{1}{c}{n} & \multicolumn{1}{c}{c} & \multicolumn{1}{c}{[\%]} & \multicolumn{1}{c}{[Jy]}\\ 
 \hline
\endfirsthead
\multicolumn{5}{l}{\scriptsize $^{*}$ The derived $(n,c)$ parameters are valid only when the surface brightness units of the input {\sl AKARI\/} slow-scan maps are given in MJy\,sr$^{-1}$.}
\endlastfoot
N60        & 0.91$\pm$0.01 & 1.08$\pm$0.01 & $94 \pm 10$ &  0.14 -- 320\\
WIDE-S  & 0.90$\pm$0.01 & 1.28$\pm$0.01 & $92 \pm \phantom{1}9$ &  0.10 -- 360\\
WIDE-L  & 0.92$\pm$0.01 & 1.39$\pm$0.02 & $88 \pm \phantom{1}7$ &  0.41 -- 270\\
N160      & 0.96$\pm$0.02 & 0.52$\pm$0.01 & $92 \pm 20$ &  1.7\phantom{1} -- 250\\
\hline
\end{longtable}}
\clearpage

{\small\setlength\tabcolsep{3pt}
\begin{longtable}{lccccc}
\caption{\label{FASTscaledphoto}Results of the Flux Correction for the PSF/Photometric Reference Sources}  \\
\hline
\multicolumn{2}{l}{} & \multicolumn{4}{c}{Observed Flux}\\
\cline{3-6} 
\multicolumn{2}{l}{Observation}& \multicolumn{1}{c}{N60} & \multicolumn{1}{c}{WIDE-S} & \multicolumn{1}{c}{WIDE-L} & \multicolumn{1}{c}{N160} \\ 
\cline{1-2}
Target & OBSID & \multicolumn{1}{c}{[Jy]} & \multicolumn{1}{c}{[Jy]} & \multicolumn{1}{c}{[Jy]} & \multicolumn{1}{c}{[Jy]} \\ 
\hline
\endfirsthead
\multicolumn{6}{l}{\scriptsize Missing fluxes were due to insufficient signal ($< 3 \sigma$) or data anomaly.}\\
\endlastfoot
\multicolumn{6}{c}{Stars}\\
\hline

 $\alpha$ CMa & 5110034-001 & \phantom{11}3.47\phantom{0} $\pm$ 0.20\phantom{0}   & \phantom{00}2.37\phantom{0} $\pm$ 0.03\phantom{0} & \dots  & \dots \\ 
 $\alpha$ Boo  & 5110039-001 & \phantom{0}19.9\phantom{00} $\pm$ 0.3\phantom{00}   & \phantom{0}13.1\phantom{00} $\pm$ 0.1\phantom{00} & \phantom{11}2.81\phantom{0}$\pm$ 0.04\phantom{0} & \phantom{11}2.41 $\pm$ 0.35\\ 
 $\alpha$ Tau   & 5110045-001 & \phantom{0}16.4\phantom{00} $\pm$ 0.8\phantom{00}   & \phantom{0}11.1\phantom{00} $\pm$ 0.1\phantom{00} & \phantom{11}2.77\phantom{0} $\pm$ 0.20\phantom{0} & \phantom{11}3.96 $\pm$ 0.95\\
 HD 216386     & 5110068-001 & \phantom{11}2.36\phantom{0} $\pm$ 0.18\phantom{0}   & \phantom{11}1.65\phantom{0} $\pm$ 0.19\phantom{0} & \phantom{11}0.456 $\pm$ 0.016                                    &  \dots \\ 
 HD 98118       & 5110072-001 & \phantom{11}0.490                   $\pm$ 0.142                    & \phantom{11}0.331                   $\pm$ 0.019                   & \dots &\dots \\ 
 HD 222643     & 5110075-001 & \phantom{11}0.167                   $\pm$ 0.126                    & \phantom{11}0.111                   $\pm$ 0.012 & \dots      &\dots \\
 HD 224935     & 5110070-001 & \phantom{11}2.07\phantom{0} $\pm$ 0.24\phantom{0}   & \phantom{11}1.36\phantom{0} $\pm$ 0.18\phantom{0} &\dots   & \dots \\ 
 HD 92305       & 5110092-001 & \phantom{11}1.08\phantom{00} $\pm$ 0.13\phantom{0} &\phantom{11}0.742                    $\pm$ 0.062                  & \dots   &\dots  \\ 
 \hline
\multicolumn{6}{c}{Asteroids and Planets}\\
\hline
 241 Germania & 5011065-001 & \phantom{00}8.85\phantom{0} $\pm$ \phantom{00}0.10\phantom{0} & \phantom{00}7.49\phantom{0} $\pm$ \phantom{0}0.06\phantom{0} & \phantom{00}2.63\phantom{0} $\pm$ \phantom{0}0.04\phantom{0} & \phantom{00}2.62\phantom{0} $\pm$ \phantom{0}0.13\phantom{0} \\ 
                         & 5011165-001 & \dots                                                                                                     & \phantom{00}7.19\phantom{0} $\pm$ \phantom{0}0.30\phantom{0} & \phantom{00}2.51\phantom{0} $\pm$ \phantom{0}0.05\phantom{0}& \phantom{00}2.95\phantom{0} $\pm$ \phantom{0}0.37\phantom{0} \\
 6 Hebe            & 5011066-001 & \phantom{0}24.4\phantom{00} $\pm$ \phantom{00}0.4\phantom{00} & \phantom{0}17.5\phantom{00} $\pm$ \phantom{0}0.2\phantom{00} & \phantom{00}5.59\phantom{0} $\pm$ \phantom{0}0.35\phantom{0} & \phantom{00}4.95\phantom{0} $\pm$ \phantom{0}0.29\phantom{0} \\
                         & 5011166-001 & \phantom{0}24.9\phantom{00} $\pm$ \phantom{00}0.3\phantom{00} & \phantom{0}19.7\phantom{00} $\pm$ \phantom{0}0.2\phantom{00} & \phantom{00}5.38\phantom{0} $\pm$ \phantom{0}0.05\phantom{0} & \phantom{00}5.38\phantom{0} $\pm$ \phantom{0}0.12\phantom{0} \\
 511 Davida      & 5011067-001 & \phantom{0}17.5\phantom{00} $\pm$ \phantom{00}0.3\phantom{00} & \phantom{0}13.9\phantom{00} $\pm$ \phantom{0}0.1\phantom{00} & \phantom{00}4.77\phantom{0} $\pm$ \phantom{0}0.04\phantom{0} & \phantom{00}4.36\phantom{0} $\pm$ \phantom{0}0.09\phantom{0} \\
                         & 5011167-001 & \phantom{0}17.3\phantom{00} $\pm$ \phantom{00}0.2\phantom{00} & \phantom{0}13.2\phantom{00} $\pm$ \phantom{0}0.1\phantom{00} & \phantom{00}4.71\phantom{0} $\pm$ \phantom{0}0.06\phantom{0} & \phantom{00}3.92\phantom{0} $\pm$ \phantom{0}0.18\phantom{0} \\
 2 Pallas           & 5110027-001 & \phantom{0}64.1\phantom{00} $\pm$ \phantom{00}0.8\phantom{00} & \phantom{0}46.7\phantom{00} $\pm$ \phantom{0}0.5\phantom{00} & \phantom{0}14.7\phantom{00} $\pm$ \phantom{0}0.1\phantom{00} & \phantom{00}9.36\phantom{0} $\pm$ \phantom{0}0.36\phantom{0} \\
 1 Ceres           & 5110032-001 & 309\phantom{.000} $\pm$ \phantom{00}5\phantom{.000}                   & 188\phantom{.000} $\pm$ \phantom{0}2\phantom{.000}                  & \phantom{0}69.3\phantom{00} $\pm$ \phantom{0}0.4\phantom{00} & \phantom{0}61.3\phantom{00} $\pm$ \phantom{0}0.7\phantom{00} \\
 93 Minerva      & 5110033-001 & \phantom{00}6.93\phantom{0} $\pm$ \phantom{00}0.12\phantom{0} & \phantom{00}5.07\phantom{0} $\pm$ \phantom{0}0.05\phantom{0} & \phantom{00}1.75\phantom{0} $\pm$ \phantom{0}0.02\phantom{0} & \phantom{00}1.91\phantom{0} $\pm$ \phantom{0}0.61\phantom{0} \\
 65 Cybele        & 5110038-001 & \phantom{0}18.1\phantom{00} $\pm$ \phantom{00}0.2\phantom{00} & \phantom{0}13.5\phantom{0} $\pm$ \phantom{0}0.2\phantom{00} & \phantom{00}4.66\phantom{0} $\pm$ \phantom{0}0.04\phantom{0} & \phantom{00}3.54\phantom{0} $\pm$ \phantom{0}0.10\phantom{0} \\
 4 Vesta            & 5110047-001 & 190\phantom{.000} $\pm$ \phantom{00}2\phantom{.000}                  & 148\phantom{.000} $\pm$ \phantom{0}1\phantom{.000}                  & \phantom{0}49.0\phantom{00} $\pm$ \phantom{0}0.3\phantom{00} & \phantom{0}47.0\phantom{00} $\pm$ \phantom{0}1.0\phantom{00} \\
                         & 5110046-001 & 192\phantom{.000} $\pm$ \phantom{00}3\phantom{.000}                  & 126\phantom{.000} $\pm$ \phantom{0}1\phantom{.000}                   & \phantom{0}38.3\phantom{00} $\pm$ \phantom{0}0.3\phantom{00} & \phantom{0}38.0\phantom{00} $\pm$ \phantom{0}0.5\phantom{00} \\
 52 Europa       & 5110058-001 & \phantom{0}23.1\phantom{00} $\pm$ \phantom{00}0.3\phantom{00} & \phantom{0}18.2\phantom{00} $\pm$ \phantom{0}0.3\phantom{00} & \phantom{00}5.45\phantom{0} $\pm$ \phantom{0}0.06\phantom{0} & \phantom{00}4.84\phantom{0} $\pm$ \phantom{0}0.28\phantom{0} \\
                         & 5110059-001 & \phantom{0}23.4\phantom{00} $\pm$ \phantom{00}0.3\phantom{00} & \phantom{0}17.5\phantom{00} $\pm$ \phantom{0}0.2\phantom{00} & \phantom{00}5.97\phantom{0} $\pm$ \phantom{0}0.04\phantom{0} & \phantom{00}5.57\phantom{0} $\pm$ \phantom{0}0.07\phantom{0} \\
 Neptune          & 5110066-001 & 313\phantom{.000} $\pm$ \phantom{00}5\phantom{.000}                   & 359\phantom{.000} $\pm$ \phantom{0}4\phantom{.000}                  & 242\phantom{.000} $\pm$ \phantom{0}2\phantom{.000}                   & 252\phantom{.000} $\pm$ \phantom{0}4\phantom{.000} \\
                        & 5110067-001 & 251\phantom{.000} $\pm$ \phantom{00}3\phantom{.000}                    & 332\phantom{.000} $\pm$ \phantom{0}4\phantom{.000}                 & 178\phantom{.000} $\pm$ \phantom{0}1\phantom{.000}                    & 205\phantom{.000} $\pm$ \phantom{0}3\phantom{.000} \\
\hline
\end{longtable}}

\end{document}